\documentclass[english]{article}
\usepackage[T1]{fontenc}
\usepackage[latin9]{inputenc}
\usepackage{color}
\usepackage{float}
\usepackage{amsmath}
\usepackage{amssymb}
\usepackage{stackrel}
\usepackage{graphicx}

\makeatletter

\pdfoutput=1
\usepackage[T1]{fontenc}
\usepackage[latin9]{inputenc}
\usepackage{color}
\usepackage{float}
\usepackage{graphicx}
\usepackage{graphics}
\usepackage{esint}
\usepackage{enumerate}

\makeatletter

\usepackage{amsfonts,amssymb}
\usepackage{amsmath}
\usepackage{latexsym}
\usepackage{epsfig}
\usepackage{cite}
\usepackage{graphicx}
\usepackage[colorlinks,linkcolor=blue]{hyperref}
\newcommand{\be}{\begin{equation}}
\newcommand{\ee}{\end{equation}}

\allowdisplaybreaks 

\usepackage{babel}
\usepackage{tcolorbox}

\usepackage{tikz}

\makeatother

\usepackage{babel}
\begin{document}
{}~ \hfill\vbox{\hbox{hep-th/yymmnnn}}\break
\vskip 3.0cm
\centerline{\Large \bf Large $D$ gravity and low $D$ string via $\alpha^{\prime}$ corrections}

\vspace*{10.0ex}

\vspace*{10.0ex}
\centerline{\large Shuxuan Ying}
\vspace*{7.0ex}
\vspace*{4.0ex}

\centerline{\large \it Department of Physics}
\centerline{\large \it Chongqing University}
\centerline{\large \it Chongqing, 401331, China} \vspace*{1.0ex}

\vspace*{4.0ex}
\centerline{\large  ysxuan@cqu.edu.cn}

\vspace*{4.0ex}

\centerline{\bf Abstract} \bigskip \smallskip

In this paper, we generalize the correspondence between large $D$ gravity and low $D$ string theory to the most general case, including its T-dual solutions. It is well-known that the large  $D$ limit of the Schwarzschild-Tangherlini black hole in gravity becomes a two-dimensional near-horizon geometry. Similarly, the large $D$ limit of its T-dual solution, obtained by the Buscher rules, namely the string black hole with a naked singularity, reduces to a two-dimensional near-singularity geometry. Both of these geometries are described by the two-dimensional low-energy effective action of string theory and are related to each other by scale-factor duality. Secondly, we demonstrate that these near-horizon/singuglarity geometries, including complete  $\alpha^{\prime}$  corrections, can be described by the two-dimensional Hohm-Zwiebach action. This approach allows for the derivation of non-perturbative and non-singular solutions. Furthermore, the Hohm-Zwiebach action provides a systematic way to investigate the $\alpha^{\prime}$-corrected near-horizon/singularity geometries of different kinds of black holes, which are difficult to achieve through the Wess-Zumino-Witten (WZW) model method.

\vfill \eject
\baselineskip=16pt
\vspace*{10.0ex}
\tableofcontents

\section{Introduction}

In a recent work, it was found that black hole physics can be drastically
simplified in the large $D$ limit while keeps essential features
of $D=4$ in Einstein's gravity \cite{Soda:1993xc,Emparan:2013xia}.
This simplification can be understood from both the perspective of
the metric and the action simultaneously. From the metric viewpoint,
all neutral and non-extremal black hole exhibit universal near-horizon
geometries in the limit $D\rightarrow\infty$ \cite{Emparan:2013xia},
which can be described by a well-known two-dimensional string black
hole \cite{Witten:1991yr,Horne:1991gn}. On the other hand, considering
the Einstein-Hilbert action, as $D\rightarrow\infty$ and utilizing
the spherically symmetric black hole ansatz, the $D$-dimensional
Einstein-Hilbert action reduces to the two-dimensional bosonic string
effective action. Therefore, these two perspectives are in agreement
with each other. Futhermore, since string theory enters the story
around the near-horizon geometry of black holes at large $D$, it
is natural to investigate how $\alpha^{\prime}$ corrections in string
theory affect this region and modify the thermodynamics. So far, there
are two main approaches are used to study $\alpha^{\prime}$ corrections
in string theory: worldsheet CFT and $\alpha^{\prime}$-corrected
low-energy effective action. Then, it is interesting to explore large
$D$ gravity and low $D$ string through these two types of $\alpha^{\prime}$
corrections. 

From the viewpoint of worldsheet CFT, Witten obtained the exact metric
describing the region outside the event horizon of the two-dimensional
string black hole through the $SL\left(2,R\right)_{k}/U\left(1\right)$
gauged WZW model \cite{Witten:1991yr}. This result is conformally
invariant to all orders in $1/k$ as $k\rightarrow\infty$ (where
$k\sim1/\alpha^{\prime}$ is the Kac--Moody level). Subsequently,
Dijkgraaf, Verlinde, and Verlinde discovered the exact two-dimensional
string black hole solution for general $k$ in this model, supposed
to be correct to all orders in $\alpha^{\prime}$ \cite{Dijkgraaf:1991ba}.
Based on this work, Perry and Teo \cite{Perry:1993ry}, and Yi \cite{Yi:1993gh}
investigated its maximally extended spacetime \cite{Bars:1992sr}.
Recent studies argue that this exact two-dimensional string black
hole solution can capture the near-horizon geometry of the Schwarzschild-Tangherlini
black hole in the large $D$ limit with higer-order $\alpha^{\prime}$
corrections \cite{Chen:2021emg,Chen:2021qrz}. In other words, the
near-horizon geometry can be described by the $SL\left(2,R\right)_{k}/U\left(1\right)$
gauged WZW model. Specifically, the ratio $r_{0}/D$ (where $r_{0}$
denotes the black hole horizon and $D\rightarrow\infty$) relates
to the level $k$ of the WZW model through $k=\left(2r_{0}/D\right)^{2}$.
Since, $r_{0}/D$ needs to be finite, the solution is valid for all
orders in $\alpha^{\prime}$ due to $k\sim1/\alpha^{\prime}$. Recent
developments can be found in references \cite{Nair:2021lxz,Chen:2021dsw,Halder:2024gwe}. 

The second approach including $\alpha^{\prime}$ corrections based
on the recent remarkable progress in classifying all orders of $\alpha^{\prime}$
corrections by Hohm and Zwiebach \cite{Hohm:2019ccp,Hohm:2019jgu}.
In the 1990s, Meissner and Veneziano discovered that if all closed
string fields depend only on the time coordinate, the string effective
action explicitly exhibits $O\left(d,d\right)$ symmetry up to the
first order in $\alpha^{\prime}$ correction. This symmetry is manifested
by using the specific $O\left(d,d\right)$ matrix \cite{Meissner:1996sa,Veneziano:1991ek}.
Sen proved this to be true for all orders of $\alpha^{\prime}$ corrections,
demonstrating that if the fields are independent of $m$ coordinates,
the action possesses $O\left(m,m\right)$ symmetry \cite{Sen:1991zi,Sen:1991cn}.
This was also confirmed in references \cite{Meissner:1991zj} from
the perspective of the sigma model expansion. Note that if all the
fields only depend on one coordinate, the string effective action
possesses a \textquotedblleft scale-factor duality\textquotedblright{}
\cite{Veneziano:1991ek,Sen:1991zi}. Therefore, it is reasonable to
assume that the specific $O\left(d,d\right)$ matrix applies to all
orders of $\alpha^{\prime}$ corrections via the suitable field redefinitions
\cite{Hohm:2019ccp,Hohm:2019jgu}. Then, Hohm and Zwiebach showed
that for cosmological time-dependent backgrounds, all orders $\alpha^{\prime}$
corrections can be rewritten into simple patterns by using the specific
$O\left(d,d\right)$ matrix, and there are only first-order time derivatives
of the dilaton and graviton. This feature implies that the Hohm-Zwiebach
action can be exactly solved, and therefore it is possible to study
the non-perturbative string cosmology and black holes via complete
$\alpha^{\prime}$ corrections. Recently, the regular cosmological
solutions of Hohm-Zwiebach action are investigated in \cite{Wang:2019kez,Wang:2019dcj,Wang:2019mwi,Gasperini:2023tus,Conzinu:2023fth},
the regular string black hole solutions are studied in \cite{Ying:2022xaj,Ying:2022cix,Codina:2023fhy,Codina:2023nwz},
and the resolution of the singularity of the Schwarzschild black hole
\cite{Wu:2024eci}.

\begin{figure}[H]
\noindent \begin{centering}
\includegraphics[scale=0.75]{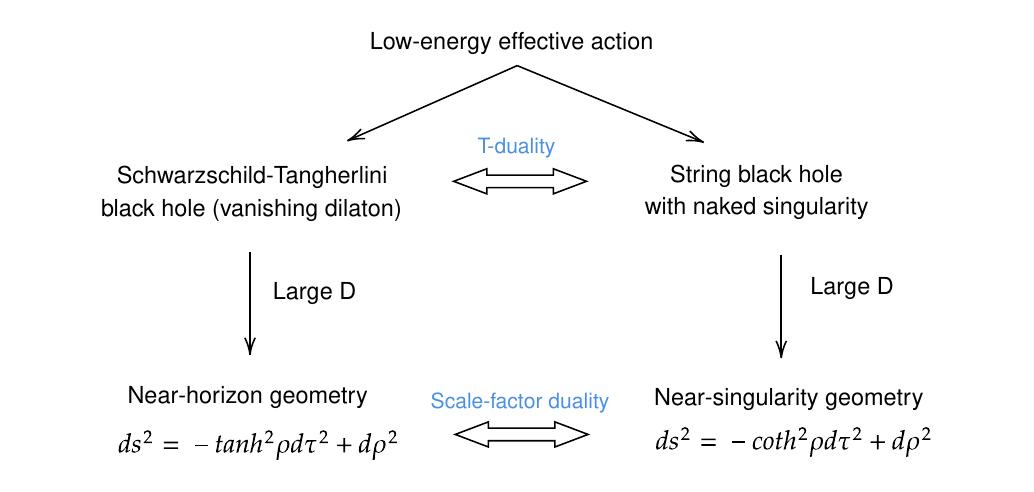}
\par\end{centering}
\caption{\label{fig:scale-factor duality}T-dual black hole solutions in string
theory reduce to two-dimensional string geometries that are related
to each other by scale-factor duality. }
\end{figure}

In this paper, we have two main aims:

\textbf{1. Large $D$ limit of the Schwarzschild-Tangherlini black
hole duals to large $D$ limit of the string black hole:} In this
part, we begin with the low-energy effective action of string theory
and calculate its higher-dimensional black hole solution. Setting
the dilaton to be constant transforms it into the Schwarzschild-Tangherlini
black hole. Its T-dual solution, describing a string black hole with
a naked singularity, is easily obtained using the Buscher rules. In
the large $D$ limit, the Schwarzschild-Tangherlini black hole reduces
to a two-dimensional near-horizon geometry, while the string black
hole reduces to a two-dimensional near-singularity geometry. Both
of these geometries are described by the two-dimensional low-energy
effective action of string theory and are related to each other by
scale-factor duality, see Figure \ref{fig:scale-factor duality}.

\begin{figure}[H]
\noindent \begin{centering}
\includegraphics[scale=0.72]{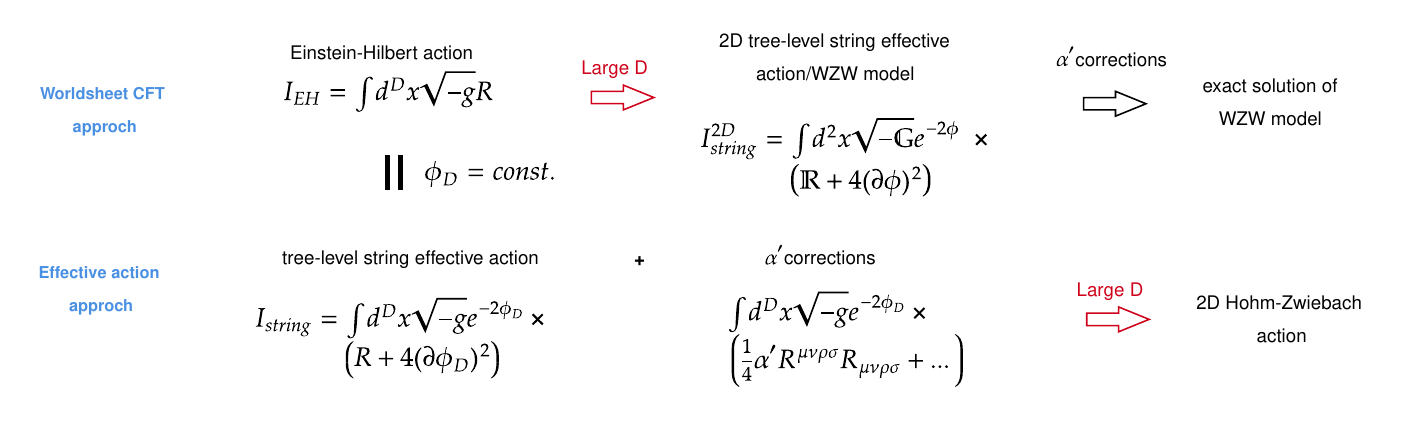}
\par\end{centering}
\caption{\label{fig:diff}The difference between two approaches. Both approaches
utilize the same ansatz (\ref{eq:ansatz}), but they take the large
$D$ limit at different stages. Note $\mathbb{R}$ is the Ricci scalar
of two-dimensional metric $\mathbb{G}_{\mu\nu}$.}
\end{figure}

\textbf{2. Investigate the complete $\alpha^{\prime}$ corrections
to previous near-horizon/singularity geometries of T-dual black holes
at large $D$:} Our objective is to demonstrate that the effects of
$\alpha^{\prime}$ corrections on these geometries can be described
by the Hohm-Zwiebach action. The difference between our approach and
the worldsheet CFT approach is illustrated in Figure \ref{fig:diff}:
In the worldsheet CFT approach, the large $D$ limit is initially
applied to the Einstein-Hilbert action, leading to the reduction of
the action to the two-dimensional string black hole, described by
the $SL\left(2,R\right)_{k}/U\left(1\right)$ gauged WZW model. By
considering the exact solution of the WZW model, the effect of $\alpha^{\prime}$
corrections can be included. On the other hand, our approach begins
with the low-energy effective action (the Einstein-Hilbert action
being a special case when considering a constant dilaton). When the
complete $\alpha^{\prime}$ corrections are included and the large
$D$ limit is taken, the result is the two-dimensional Hohm-Zwiebach
action. Note that, while the critical dimension of bosonic string
theory is 26, the relative error between 26 dimensions and the large
$D$ approximation is not significant enough to change the physics
\cite{Chen:2021emg}.

\vspace*{2.0ex}

To illustrate this result, we consider the $D$-dimensional low-energy
effective action with undetermined complete $\alpha^{\prime}$ corrections:

\begin{equation}
I_{String}=\frac{1}{16\pi G_{D}}\int d^{D}x\sqrt{-g}e^{-2\phi_{D}}\left(R+4\left(\partial\phi_{D}\right)^{2}+\frac{1}{4}\alpha^{\prime}R^{\mu\nu\rho\sigma}R_{\mu\nu\rho\sigma}+\alpha^{\prime2}\left(\ldots\right)+\ldots\right).\label{eq:action}
\end{equation}

\noindent We also utilize the ansatz as follows:

\begin{equation}
ds^{2}=\underset{2\;\mathrm{dimensions}}{\underbrace{\mathbb{G}_{\mu\nu}\left(x^{\mu}\right)dx^{\mu}dx^{\nu}}}+\underset{n+1\;\mathrm{dimensional\;sphere}}{\underbrace{r^{2}d\Omega_{n+1}^{2}},}\quad r^{2}\equiv r_{0}^{2}e^{-4\phi\left(x^{\mu}\right)/\left(n+1\right)},\label{eq:ansatz}
\end{equation}

\noindent where $\mathbb{G}_{\mu\nu}$ is the two-dimensional metric,
$r_{0}$ denotes the event horizon, and $\phi\left(x^{\mu}\right)$
plays a role as a scalar field in the reduced action. As $D\rightarrow\infty$,
the ansatz (\ref{eq:ansatz}) only depends on one coordinate, implying
that all closed string fields $g$ and $\phi_{D}$ of the action (\ref{eq:action})
only depend on one coordinate. Therefore, Based on Sen's proof \cite{Sen:1991zi,Sen:1991cn},
the $O\left(d,d\right)$ symmetry emerges in the action (\ref{eq:action})
at the leading order of $\frac{1}{D}$. Moreover, since the radius
of the metric above is fixed to the constant $r_{0}$, the spherical
coordinates can be integrated out. It implies that the action can
be reduced to two dimensions, and the $O\left(d,d\right)$ symmetry
becomes $O\left(1,1\right)$ symmetry. Then, using the same derivation
as Hohm and Zwiebach, we can obtain the two-dimensional $O\left(1,1\right)$
invariant effective action with complete $\alpha^{\prime}$ corrections.
Based on this result, we compute the non-perturbative and non-singular
two-dimensional geometries, including complete $\alpha^{\prime}$
corrections, that describe the near-horizon/singularity geometries
of large $D$ black holes derived from gravity and string theory.
Finally, due to the advantages of the effective action approach, we
provide some examples where the Hohm-Zwiebach action can be directly
applied.

This paper is organized as follows. In Section 2, We provide a brief
review of scale-factor duality and the shifted dilaton in string theory.
In section 3, we calculate the higher-dimensional black hole solution
and its T-dual in string theory. Subsequently, we demonstrate the
scale-factor duality between these two black hole solutions in the
large $D$ limit. In Section 4, we show that the complete $\alpha^{\prime}$
correction of the near-horizon/singularity geometries can be described
by the Hohm-Zwiebach action. Next, we compute the $\alpha^{\prime}$-corrected
near-horizon/singularity geometries of black holes using the Hohm-Zwiebach
action. Section 5 presents several types of black holes in the large
$D$ limit, which can be applied to the Hohm-Zwiebach action to include
complete $\alpha^{\prime}$ corrections directly. The final section
contains a discussion and conclusion.

\section{Scale-factor duality and shifted dilaton in string theory}

In this section, we plan to review scale-factor duality and the shifted
dilaton in the low-energy effective action of string theory. When
all fields depend only on coordinates, the low-energy effective action
of string theory exhibits \textquotedbl scale-factor duality\textquotedbl{}
\cite{Veneziano:1991ek,Sen:1991zi}, which is a specific manifestation
of the $O\left(d,d\right)$ symmetry. A comprehensive review of this
concept can be found in \cite{string}.

We start with the tree-level low-energy effective action, focusing
only on the gravitational sector without any matter sources. The anti-symmetric
Kalb-Ramond field $b_{ij}$ is set to vanish. The two-dimensional
action is given by:

\begin{equation}
I_{String}=\frac{1}{16\pi G_{2}}\int d^{2}x\sqrt{-g}e^{-2\phi}\left(R+4\left(\partial\phi\right)^{2}+4\lambda^{2}\right),
\end{equation}

\noindent where $\phi$ represents the dilaton, $g_{\mu\nu}$ is the
string metric and $\lambda$ is a constant related to the cosmological
constant. The EOM derived from this action are:

\begin{eqnarray}
R+4\nabla^{2}\phi-4\left(\partial_{\mu}\phi\right)^{2}+4\lambda^{2} & = & 0,\nonumber \\
R_{\mu\nu}+2\nabla_{\mu}\nabla_{\nu}\phi & = & 0.\label{eq:EOM for dila gravi}
\end{eqnarray}

\noindent Considering the domain-wall ansatz:

\begin{equation}
ds^{2}=-a\left(\rho\right)^{2}dt^{2}+b\left(\rho\right)^{2}d\rho^{2},
\end{equation}

\noindent the EOM (\ref{eq:EOM for dila gravi}) for the graviton
and the dilaton take the form:

\begin{eqnarray}
H\mathcal{D}\phi-\left(\mathcal{D}\phi\right)^{2}+\lambda^{2} & = & 0,\nonumber \\
2\mathcal{D}^{2}\phi-\left(\mathcal{D}H\right)-H^{2} & = & 0,\nonumber \\
\left(\mathcal{D}H\right)+H^{2}-2H\left(\mathcal{D}\phi\right) & = & 0,\label{eq:EOM for metric}
\end{eqnarray}

\noindent where $\mathcal{D}=\frac{1}{b\left(\rho\right)}\partial_{\rho}$
and $H\left(r\right)\equiv\frac{\mathcal{D}a\left(\rho\right)}{a\left(\rho\right)}$.
These EOM are invariant under the transformations known as \textquotedbl scale-factor
duality,\textquotedbl{} given by:

\begin{equation}
a\left(\rho\right)\rightarrow\tilde{a}\left(\rho\right)=\frac{1}{a\left(\rho\right)},\qquad\phi\left(\rho\right)\rightarrow\tilde{\phi}\left(\rho\right)=\phi\left(\rho\right)-\frac{1}{2}\ln\left(-g_{00}\right),
\end{equation}

\noindent where $\tilde{\phi}$ is called the \textquotedbl shifted
dilaton\textquotedbl . Under these transformations, the EOM (\ref{eq:EOM for metric})
for the metric become:

\begin{eqnarray}
\tilde{H}\mathcal{D}\tilde{\phi}-\left(\mathcal{D}\tilde{\phi}\right)^{2}+\lambda^{2} & = & 0,\nonumber \\
2\mathcal{D}^{2}\tilde{\phi}-\left(\mathcal{D}\tilde{H}\right)-\tilde{H}^{2} & = & 0,\nonumber \\
\left(\mathcal{D}\tilde{H}\right)+\tilde{H}^{2}-2\tilde{H}\left(\mathcal{D}\tilde{\phi}\right) & = & 0,\label{eq:EOM for metric dual}
\end{eqnarray}

\noindent where $\tilde{H}\equiv\frac{\mathcal{D}\tilde{a}\left(\rho\right)}{\tilde{a}\left(\rho\right)}$.
The solutions to these EOM (\ref{eq:EOM for metric}) and their duals
(\ref{eq:EOM for metric dual}) are given by:

\begin{eqnarray}
 &  & a\left(\rho\right)=\tanh\left(\lambda\rho\right),\qquad b\left(\rho\right)=1,\qquad\phi\left(\rho\right)=-\frac{1}{2}\ln\cosh^{2}\left(\lambda\rho\right),\nonumber \\
 &  & \tilde{a}\left(\rho\right)=\coth\left(\lambda\rho\right),\qquad\tilde{b}\left(\rho\right)=1,\qquad\tilde{\phi}\left(\rho\right)=-\frac{1}{2}\ln\sinh^{2}\left(\lambda\rho\right),
\end{eqnarray}

\noindent where $\lambda$ can also be rescaled into $\rho$.

\section{Large $D$ black holes in gravity and string theory}

In this section, we first calculate the $D$-dimensional black hole
solution of the tree-level low-energy effective action of string theory.
We then apply the Buscher rules to this metric to obtain its T-dual
version. After setting the parameters simultaneously in these two
dual metrics, one metric reduces to the Schwarzschild-Tangherlini
black hole in gravity, and the other becomes the string black hole
with a naked singularity. Note that the T-duality between these two
metrics still exists. Considering the large $D$ limit of these two
metrics, the Schwarzschild-Tangherlini black hole becomes a two-dimensional
near-horizon geometry that can be described by the string action,
while the naked singularity string black hole reduces to a two-dimensional
near-singularity geometry. The ordinary T-duality between these two
geometries reduces to scale-factor duality in a two-dimensional background
at large $D$. In both cases, we will derive these two-dimensional
geometries independently through both metric and action descriptions.

We begin with the low-energy effective action of string theory in
$D$ dimensions, assuming a vanishing Kalb-Ramond field and cosmological
constant:

\begin{equation}
I_{\mathrm{String}}=\frac{1}{16\pi G_{D}}\int d^{D}x\sqrt{-g}e^{-2\phi}\left(R+4\left(\partial\phi\right)^{2}\right),
\end{equation}

\noindent where $\phi$ represents the dilaton field. To extend the
four-dimensional black hole solution \cite{Kar:1998rv} to higher
dimensions and its T-dual, we can proceed as follows:

\vspace*{2.0ex}

\noindent \textbf{Ordinary solution:}

\noindent The ordinary higher-dimensional black hole solution is given
by:

\begin{equation}
ds_{\mathrm{Ordinary}}^{2}=-\left(1-\left(\frac{2\eta}{r}\right)^{n}\right)^{\frac{m+\sigma}{\eta}}dt^{2}+\left(1-\left(\frac{2\eta}{r}\right)^{n}\right)^{\frac{\sigma-m}{\eta}}dr^{2}+\left(1-\left(\frac{2\eta}{r}\right)^{n}\right)^{1+\frac{\sigma-m}{\eta}}r^{2}d\Omega_{n+1}^{2},\label{eq:D string black hole}
\end{equation}

\noindent where $d\Omega_{n+1}^{2}$ denotes the metric of an $\left(n+1\right)$-dimensional
sphere, and the dilaton solution is:

\begin{equation}
\phi_{\mathrm{Ordinary}}\left(r\right)=\frac{1}{4\eta}\left(\left(n-1\right)\left(\eta-m\right)+\left(n+1\right)\sigma\right)\ln\left(1-\left(\frac{2\eta}{r}\right)^{n}\right),
\end{equation}

\noindent subject to the constraint:

\begin{equation}
-\left(n+1\right)m^{2}-\left(n-3\right)\eta^{2}-\left(n+1\right)\sigma^{2}+2\left(n-1\right)m\left(\eta+\sigma\right)-2\left(n-1\right)\eta\sigma=0.\label{eq:constraint}
\end{equation}

\vspace*{2.0ex}

\noindent \textbf{T-dual solution:}

\noindent Applying the Buscher rules with a vanishing Kalb-Ramond
field along the $g_{00}$ direction,

\begin{equation}
\tilde{g}_{00}=\frac{1}{g_{00}},\qquad\phi_{\mathrm{T-dual}}=\phi_{\mathrm{Ordinary}}-\frac{1}{2}\ln\left(-g_{00}\right),
\end{equation}

\noindent the T-dual black hole solution is:

\begin{equation}
ds_{\mathrm{T-dual}}^{2}=-\left(1-\left(\frac{2\eta}{r}\right)^{n}\right)^{\frac{-m-\sigma}{\eta}}dt^{2}+\left(1-\left(\frac{2\eta}{r}\right)^{n}\right)^{\frac{\sigma-m}{\eta}}dr^{2}+\left(1-\left(\frac{2\eta}{r}\right)^{n}\right)^{1+\frac{\sigma-m}{\eta}}r^{2}d\Omega_{n+1}^{2},\label{eq:D string black hole dual}
\end{equation}

\noindent with the shifted dilaton solution:

\begin{equation}
\phi_{\mathrm{T-dual}}\left(r\right)=\frac{1}{4\eta}\left(\left(n-1\right)\left(\eta+\sigma\right)-\left(n+1\right)m\right)\ln\left(1-\left(\frac{2\eta}{r}\right)^{n}\right).
\end{equation}

\noindent The constraint for the parameters is same as (\ref{eq:constraint}).

\vspace*{2.0ex}

To maintain spherical symmetry in both metrics, we choose identical
values for $m$, $\eta$, and $\sigma$, such that $\sigma=0$ and
$m=\eta$. Consequently, the metrics (\ref{eq:D string black hole})
and (\ref{eq:D string black hole dual}) are given by:

\begin{eqnarray}
ds_{\mathrm{Ordinary}}^{2}=-\left(1-\left(\frac{r_{0}}{r}\right)^{n}\right)dt^{2}+\frac{dr^{2}}{\left(1-\left(\frac{r_{0}}{r}\right)^{n}\right)}+r^{2}d\Omega_{n+1}^{2}, &  & \phi_{\mathrm{Ordinary}}\left(r\right)=0,\label{eq:ordinary metric}\\
ds_{\mathrm{T-dual}}^{2}=-\left(1-\left(\frac{r_{0}}{r}\right)^{n}\right)^{-1}dt^{2}+\frac{dr^{2}}{\left(1-\left(\frac{r_{0}}{r}\right)^{n}\right)}+r^{2}d\Omega_{n+1}^{2}, &  & \phi_{\mathrm{T-dual}}\left(r\right)=-\frac{1}{2}\ln\left(1-\left(\frac{r_{0}}{r}\right)^{n}\right),\label{eq:T dual metric}
\end{eqnarray}

\noindent where $r_{0}=2m$. Here, $ds_{\mathrm{Ordinary}}^{2}$ represents
the Schwarzschild-Tangherlini black hole solution of Einstein's gravity
with a vanishing dilaton, where $r_{0}$ denotes the event horizon.
On the other hand, $ds_{\mathrm{T-dual}}^{2}$ corresponds to the
string black hole solution with a non-trivial dilaton, where $r_{0}$
indicates the curvature singularity.

Next, we investigate the Large $D$ limit of these dual black hole
solutions.

\subsection{Large $D$ black hole (ordinary solution) in gravity}

The large $D$ Schwarzschild-Tangherlini black hole in Einstein's
gravity has been extensively examined in reference \cite{Emparan:2013xia}.
It can be approached through two perspectives: metric and action descriptions.
A concise overview is provided below:

\vspace*{2.0ex}

\noindent \textbf{Metric description:}

\noindent We begin with the Schwarzschild-Tangherlini solution (\ref{eq:ordinary metric})
in $D=3+n$ dimensions that we obtained before:

\begin{equation}
ds_{\mathrm{Ordinary}}^{2}=-\left(1-\left(\frac{r_{0}}{r}\right)^{n}\right)dt^{2}+\frac{dr^{2}}{\left(1-\left(\frac{r_{0}}{r}\right)^{n}\right)}+r^{2}d\Omega_{n+1}^{2},\quad\phi_{\mathrm{Ordinary}}\left(r\right)=0.\label{eq:ST black hole}
\end{equation}

\noindent If $r>r_{0}$, in the limit $n\rightarrow\infty$, the metric
becomes flat, and this is also known as the far region. To study the
near region, we can use the coordinate transformation $\mathrm{R}=\left(\frac{r}{r_{0}}\right)^{n}$,
the metric (\ref{eq:ST black hole}) becomes

\begin{equation}
ds_{\mathrm{Ordinary}}^{2}=-\frac{\mathrm{R}-1}{\mathrm{R}}dt^{2}+\frac{r_{0}^{2}}{n^{2}}\mathrm{R}^{\frac{2}{n}}\frac{1}{\mathrm{R}\left(\mathrm{R}-1\right)}d\mathrm{R}^{2}+r_{0}^{2}\mathrm{R}^{\frac{2}{n}}d\Omega_{n+1}^{2}.\label{eq:metric 1}
\end{equation}

\noindent The near-horizon metric can be obtained by requiring $\ln\mathrm{R}\ll n$,
which gives:

\begin{equation}
ds_{\mathrm{Ordinary}}^{2}=-\frac{\mathrm{R}-1}{\mathrm{R}}dt^{2}+\frac{r_{0}^{2}}{n^{2}}\frac{1}{\mathrm{R}\left(\mathrm{R}-1\right)}d\mathrm{R}^{2}+r_{0}^{2}d\Omega_{n+1}^{2}.
\end{equation}

\noindent Since $r^{2}d\Omega_{n+1}^{2}$ becomes $r_{0}^{2}d\Omega_{n+1}^{2}$,
the metric then describes the near-horizon geometry. We can further
study this region by using the following coordinate transformations:

\begin{equation}
\mathrm{R}=\cosh^{2}\rho,\qquad d\tau=\frac{n}{2r_{0}}dt,
\end{equation}

\noindent the metric becomes

\begin{equation}
ds_{\mathrm{Ordinary}}^{2}=\left(\frac{2r_{0}}{n}\right)^{2}\left(-\tanh^{2}\rho d\tau^{2}+d\rho^{2}\right)+r_{0}^{2}d\Omega_{n+1},
\end{equation}

\noindent The first part of the metric describes the near-horizon
geometry of the Schwarzschild-Tangherlini black hole in the large
$D$ limit, which corresponds to the two-dimensional string black
hole. This result holds for all neutral and non-extremal black holes.

\vspace*{2.0ex}

\noindent \textbf{Action description:}

\noindent Alternatively, the metric description above is equivalent
to the action description. Let us recall the low-energy effective
action with vanishing dilaton field, the action reduces to the Hilbert-Einstein
action:

\begin{equation}
I_{\mathrm{String}}=\frac{1}{16\pi G_{D}}\int d^{D}x\sqrt{-g}R,
\end{equation}

\noindent and utilizing the dimensional reduction on a sphere to introduce
the dilaton field $\phi\left(x^{\mu}\right)$

\begin{equation}
ds^{2}=\underset{2\;\mathrm{dimensions}}{\underbrace{\mathbb{G}_{\mu\nu}\left(x^{\mu}\right)dx^{\mu}dx^{\nu}}}+\underset{n+1\;\mathrm{dimensional\;sphere}}{\underbrace{r_{0}^{2}e^{-4\phi\left(x^{\mu}\right)/\left(n+1\right)}d\Omega_{n+1}^{2}}},\label{eq:metric 2}
\end{equation}

\noindent the Einstein-Hilbert action becomes

\begin{equation}
I_{\mathrm{String}}=\frac{\Omega_{n+1}r_{0}^{n+1}}{16\pi G_{D}}\int d^{2}x\sqrt{-\mathbb{G}}e^{-2\phi}\left(\mathbb{R}+\frac{4n}{n+1}\left(\partial\phi\right)^{2}+\frac{n\left(n+1\right)}{r_{0}^{2}}e^{-4\phi/\left(n+1\right)}\right),
\end{equation}

\noindent where $\mathbb{R}$ is the Ricci scalar of two-dimensional
metric $\mathbb{G}_{\mu\nu}$, $\phi$ can be seen as the two-dimensional
dilaton, and a volume of unit sphere is given by $\Omega_{n+1}=2\pi^{\frac{n+2}{2}}/\Gamma\left(\frac{n+2}{2}\right)$.
When $n\rightarrow\infty$, the action reduces to the two-dimensional
string effective action. This action possesses an $SU\left(2\right)_{k}/U\left(1\right)$
symmetry:

\begin{equation}
I_{\mathrm{String}}=\frac{1}{16\pi G_{2}}\int d^{2}x\sqrt{-\mathbb{G}}e^{-2\phi}\left(\mathbb{R}+4\left(\partial\phi\right)^{2}+4\lambda^{2}\right),
\end{equation}

\noindent where $G_{2}=\underset{n\rightarrow\infty}{\lim}\frac{G_{D}}{\Omega_{n+1}r_{0}^{n+1}}$
and $\lambda=\frac{n}{2r_{0}}$.\textcolor{red}{{} }To determine the
dilaton solution, we analyze the metrics (\ref{eq:metric 1}) and
(\ref{eq:metric 2}), yielding:

\begin{equation}
\phi\left(\mathrm{R}\right)=-\frac{n+1}{2n}\ln\mathrm{R}\overset{n\rightarrow\infty}{\longrightarrow}-\frac{1}{2}\ln\mathrm{R}.
\end{equation}

\noindent By applying the coordinate transformation $\mathrm{R}=\cosh^{2}\rho$,
we derive the final solutions for the two-dimensional low-energy effective
action:

\begin{equation}
ds_{\mathrm{Ordinary}}^{2}=\left(\frac{2r_{0}}{n}\right)^{2}\left(-\tanh^{2}\rho d\tau^{2}+d\rho^{2}\right),\qquad\phi\left(\rho\right)=-\frac{1}{2}\ln\cosh^{2}\rho.
\end{equation}

\noindent Moreover, note the string length corresponds to $\ell_{s}=\sqrt{\alpha^{\prime}}\sim\frac{r_{0}}{D}$,
which also indicates the length scale away from the event horizon,
namely $r-r_{0}\sim\frac{r_{0}}{D}$. When $\lambda\rightarrow\infty$,
$\alpha^{\prime}\rightarrow0$, and we do not need to consider the
string corrections. On the other hand, if we keep the cosmological
constant $\lambda$ finite and $r_{0}\sim D$, then all orders of
$\alpha^{\prime}$ corrections of string theory should be considered.
In the following sections, we will consider the higher curvature corrections
by assuming $r_{0}\sim D$.

\subsection{Large $D$ black hole (T-dual solution) in string theory}

\vspace*{2.0ex}

\noindent \textbf{Metric description:}

\noindent Recalling the $D=n+3$ dimensional string black hole solution
(\ref{eq:T dual metric}) of the low-energy effective action featuring
a non-trivial dilaton:

\begin{equation}
ds_{\mathrm{T-dual}}^{2}=-\left(1-\left(\frac{r_{0}}{r}\right)^{n}\right)^{-1}dt^{2}+\frac{dr^{2}}{\left(1-\left(\frac{r_{0}}{r}\right)^{n}\right)}+r^{2}d\Omega_{n+1}^{2},\quad\phi_{\mathrm{T-dual}}\left(r\right)=-\frac{1}{2}\ln\left(1-\left(\frac{r_{0}}{r}\right)^{n}\right).\label{eq:SY black hole}
\end{equation}

\noindent Applying the coordinate transformation $\mathrm{R}=\left(\frac{r}{r_{0}}\right)^{n}$,
the metric (\ref{eq:SY black hole}) transforms to:

\begin{equation}
ds^{2}=\frac{\mathrm{R}}{\mathrm{R}-1}\left(-dt^{2}+\frac{r_{0}^{2}}{n^{2}}\mathrm{R^{\frac{2-2n}{n}}}d\mathrm{R}^{2}\right)+r_{0}^{2}\mathrm{R}^{\frac{2}{n}}d\Omega_{n+1}^{2}.\label{eq:SY specific metric}
\end{equation}

\noindent The curvature singularity locates at $r=r_{0}$, corresponding
to $\mathrm{R}=1$. As $n\rightarrow\infty$, the $D$-dimensional
metric (\ref{eq:SY black hole}) simplifies to the two-dimensional
near-singularity geometry:

\begin{equation}
ds^{2}=-\frac{\mathrm{R}}{\mathrm{R}-1}dt^{2}+\frac{r_{0}^{2}}{n^{2}}\frac{1}{\mathrm{R}\left(\mathrm{R}-1\right)}d\mathrm{R}^{2}+r_{0}^{2}d\Omega_{n+1}^{2}.\label{eq:2D metric in string}
\end{equation}

\noindent Further employing the coordinate transformations:

\begin{equation}
\mathrm{R}=\cosh^{2}\rho,\qquad d\tau=\frac{n}{2r_{0}}dt,
\end{equation}

\noindent the metric becomes:

\begin{equation}
ds^{2}=\left(\frac{2r_{0}}{n}\right)^{2}\left(-\coth^{2}\rho d\tau^{2}+d\rho^{2}\right)+r_{0}^{2}d\Omega_{n+1},
\end{equation}

\noindent In this limit, the dilaton field (\ref{eq:SY black hole})
is given by:

\begin{equation}
\phi_{\mathrm{T-dual}}\left(\rho\right)=-\frac{1}{2}\ln\tanh^{2}\rho.
\end{equation}

\noindent Note that $\phi_{\mathrm{T-dual}}\left(\rho\right)$ does
not directly satisfy the reduced two-dimensional low-energy effective
action due to the spherical reduction introducing the extra scalar
field $\phi\left(\rho\right)$. Understanding this result requires
consideration of the following action description.

\vspace*{2.0ex}

\noindent \textbf{Action description:}

\noindent Let us recall the tree-level low-energy effective action:

\begin{equation}
I_{\mathrm{String}}=\frac{1}{16\pi G_{D}}\int d^{D}x\sqrt{-g}e^{-2\phi_{\mathrm{T-dual}}}\left(R+4\left(\partial\phi_{\mathrm{T-dual}}\right)^{2}\right).\label{eq:}
\end{equation}

\noindent To investigate the behavior of this action in the large
$D$ limit, we also perform dimensional reduction on a sphere:

\begin{equation}
ds^{2}=\underset{2\;\mathrm{dimensions}}{\underbrace{\mathbb{G}_{\mu\nu}dx^{\mu}dx^{\nu}}}+\underset{n+1\;\mathrm{dimensional\;sphere}}{\underbrace{r_{0}^{2}e^{-4\phi\left(x\right)/\left(n+1\right)}d\Omega_{n+1}^{2}}}.
\end{equation}

\noindent Comparing this with the specific metric (\ref{eq:SY specific metric}),
as $n\rightarrow\infty$, we find:

\begin{equation}
r_{0}^{2}e^{-4\phi\left(x\right)/\left(n+1\right)}=r_{0}^{2}\mathrm{R}^{\frac{2}{n}}\qquad\Rightarrow\qquad\phi\left(\mathrm{R}\right)=-\frac{1}{2}\ln\mathrm{R}.
\end{equation}

\noindent Under the coordinate transformation $\mathrm{R}=\cosh^{2}\rho$,
this gives:

\begin{equation}
\phi\left(\rho\right)=-\frac{1}{2}\ln\cosh^{2}\rho.
\end{equation}

\noindent Using this dimensional reduction, the action (\ref{eq:})
becomes:
\begin{eqnarray}
I_{\mathrm{String}} & = & \frac{\Omega_{n+1}r_{0}^{n+1}}{16\pi G_{D}}\int d^{2}x\sqrt{-\mathbb{G}}e^{-2\left(\phi_{\mathrm{T-dual}}+\phi\right)}\left[\mathbb{R}+4\left(\partial\phi_{\mathrm{T-dual}}\right)^{2}\right.\nonumber \\
 &  & \left.+8\partial\phi\partial\phi_{\mathrm{T-dual}}+\frac{4n}{n+1}\left(\partial\phi\right)^{2}+\frac{n\left(n+1\right)}{r_{0}^{2}}e^{\frac{4\phi}{n+1}}\right],
\end{eqnarray}

\noindent where $\mathbb{R}$ is the Ricci scalar of the two-dimensional
metric $\mathbb{G}_{\mu\nu}$. In the limit $n\rightarrow\infty$,
he action reduces to the two-dimensional low-energy effective action:

\begin{equation}
I_{\mathrm{String}}=\frac{1}{16\pi G_{2}}\int d^{2}x\sqrt{-\mathbb{G}}e^{-2\tilde{\phi}}\left(\mathbb{R}+4\left(\partial\tilde{\phi}\right)^{2}+4\lambda^{2}\right),\qquad\tilde{\phi}\left(\rho\right)\equiv\phi\left(\rho\right)+\phi_{\mathrm{T-dual}}\left(\rho\right),\label{eq:2D action}
\end{equation}

\noindent where $G_{2}=\underset{n\rightarrow\infty}{\lim}\frac{G_{D}}{\Omega_{n+1}r_{0}^{n+1}}$
and $\lambda=\frac{n}{2r_{0}}$. Therefore, the final solutions in
the two-dimensional theory are:

\begin{eqnarray}
ds^{2} & = & \left(\frac{2r_{0}}{n}\right)^{2}\left(-\coth^{2}\rho d\tau^{2}+d\rho^{2}\right),\nonumber \\
\tilde{\phi}\left(\rho\right) & = & \phi\left(\rho\right)+\phi_{\mathrm{T-dual}}\left(\rho\right)=-\frac{1}{2}\ln\cosh^{2}\rho-\frac{1}{2}\ln\tanh^{2}\rho=-\frac{1}{2}\ln\sinh^{2}\rho.
\end{eqnarray}

\subsection{Scale-factor duality between gravity black hole and string black
hole}

In the previous section, we explored the large $D$ limit of the Schwarzschild-Tangherlini
black hole in gravity, where the resulting geometry signifies the
two-dimensional near-horizon geometry of the black hole. This geometry
is described by the two-dimensional low-energy effective action:

\begin{equation}
ds^{2}=\left(\frac{2r_{0}}{n}\right)^{2}\left(-\tanh^{2}\rho d\tau^{2}+d\rho^{2}\right),\qquad\phi\left(\rho\right)=-\frac{1}{2}\ln\cosh^{2}\rho.\label{eq:2d metric1}
\end{equation}

\noindent Similarly, we also considered the large $D$ limit of the
black hole in string theory, which reduces to a two-dimensional geometry
describing the region near the singularity. This geometry is also
a solution of two-dimensional string theory and is given by:

\begin{equation}
ds^{2}=\left(\frac{2r_{0}}{n}\right)^{2}\left(-\coth^{2}\rho d\tau^{2}+d\rho^{2}\right),\qquad\tilde{\phi}\left(\rho\right)=-\frac{1}{2}\ln\sinh^{2}\rho.\label{eq:2d metric2}
\end{equation}

\noindent As anticipated, the large $D$ limit maintains the T-duality
between these two solutions. Notably, these solutions exhibit scale-factor
duality in string theory, as introduced in Section 2:

\begin{equation}
\tanh\rho=\frac{1}{\coth\rho},\qquad\tilde{\phi}\left(\rho\right)=\phi\left(\rho\right)-\frac{1}{2}\ln\left(-\mathbb{G}_{00}\right),
\end{equation}

\noindent where $\tilde{\phi}\left(\rho\right)$ represents the shifted
dilaton. In essence, the large $D$ limit of the Schwarzschild-Tangherlini
black hole in gravity is dual to the large $D$ limit of the black
hole in string theory.

\section{Large $D$ black holes via $\alpha^{\prime}$ corrections}

In this section, we demonstrate that the low-energy effective action,
incorporating complete $\alpha^{\prime}$ corrections under the dimensional
reduction ansatz (\ref{eq:ansatz}), exhibits $O\left(1,1\right)$
symmetry. This symmetry implies that the action can be effectively
described by the two-dimensional Hohm-Zwiebach action. This systematic
approach facilitates the study of $\alpha^{\prime}$-corrected near-horizon/singularity
geometries of various types of black holes.

\subsection{Large $D$ string effective action with complete $\alpha^{\prime}$
corrections}

In the previous section, we studied the tree-level low-energy effective
action:

\begin{equation}
I_{\mathrm{String}}^{\left(0\right)}=\frac{1}{16\pi G_{D}}\int d^{D}x\sqrt{-g}e^{-2\phi_{D}}\left(R+4\left(\partial\phi_{D}\right)^{2}\right).\label{eq:action alpha}
\end{equation}

\noindent where $\phi_{D}\left(x\right)$ represents the $D$-dimensional
dilaton. Using the dimensional reduction on a sphere:

\begin{equation}
ds^{2}=\underset{2\;\mathrm{dimensions}}{\underbrace{\mathbb{G}_{\mu\nu}\left(x\right)dx^{\mu}dx^{\nu}}}+\underset{n+1\;\mathrm{dimensional\;sphere}}{\underbrace{r_{0}^{2}e^{-4\phi\left(x\right)/\left(n+1\right)}d\Omega_{n+1}^{2}}},\label{eq:ansatz alpha}
\end{equation}

\noindent the action (\ref{eq:action alpha}) transforms into:

\begin{equation}
I_{\mathrm{String}}^{\left(0\right)}=\Omega_{n+1}r_{0}^{n+1}\int d^{2}x\sqrt{-\mathbb{G}}e^{-2\left(\phi+\phi_{D}\right)}\left[\mathbb{R}+\frac{4n}{n+1}\left(\partial\phi\right)^{2}+8\partial\phi\partial\phi_{D}+4\left(\partial\phi_{D}\right)^{2}+\frac{n\left(n+1\right)}{r_{0}^{2}}e^{\frac{4\phi}{n+1}}\right],
\end{equation}

\noindent In the limit $n\rightarrow\infty$, this action reduces
to the two-dimensional string effective action after identifying $\lambda=\frac{n}{2r_{0}}$
and $\tilde{\phi}=\phi+\phi_{D}$:

\begin{equation}
I_{\mathrm{String}}^{\left(0\right)}=\Omega_{n+1}r_{0}^{n+1}\int d^{2}x\sqrt{-\mathbb{G}}e^{-2\tilde{\phi}}\left[\mathbb{R}+4\left(\partial\tilde{\phi}\right)^{2}+4\lambda^{2}\right]+\mathcal{O}\left(\frac{1}{n}\right),
\end{equation}

\noindent To extend this analysis to include first-order $\alpha^{\prime}$
correction, we consider the action:

\begin{equation}
I_{\mathrm{String}}^{\left(1\right)}=\frac{1}{16\pi G_{D}}\int d^{D}x\sqrt{-g}e^{-2\phi_{D}}\left(R+4\left(\partial\phi_{D}\right)^{2}+\frac{1}{4}\alpha^{\prime}R^{\mu\nu\rho\sigma}R_{\mu\nu\rho\sigma}\right),\label{eq:string alpha}
\end{equation}

\noindent Using the ansatz (\ref{eq:ansatz alpha}) and taking the
large $n$ limit, the action (\ref{eq:string alpha}) becomes:

\begin{equation}
I_{\mathrm{String}}^{\left(1\right)}=\frac{1}{16\pi G_{2}}\int d^{2}x\sqrt{-\mathbb{G}}e^{-2\tilde{\phi}}\left[\mathbb{R}+4\left(\partial\tilde{\phi}\right)^{2}+4\lambda^{2}+\frac{1}{4}\alpha^{\prime}\mathbb{R}^{\mu\nu\rho\sigma}\mathbb{R}_{\mu\nu\rho\sigma}\right]+\mathcal{O}\left(\frac{1}{n}\right),\label{eq:reduced action}
\end{equation}

\noindent where $G_{2}=\underset{n\rightarrow\infty}{\lim}\frac{G_{D}}{\Omega_{n+1}r_{0}^{n+1}}$
and $\mathbb{R}^{\mu\nu\rho\sigma}\mathbb{R}_{\mu\nu\rho\sigma}$
is the Kretschmann scalar of the two-dimensional metric $\mathbb{G}_{\mu\nu}$.
Finally, generalizing this approach to higher-order $\alpha^{\prime}$
corrections, we also consider the ansatz (\ref{eq:ansatz alpha}).
As $n\rightarrow\infty$, the metric (\ref{eq:ansatz alpha}) simplifies
to:

\begin{equation}
ds^{2}=\underset{2\;\mathrm{dimensions}}{\underbrace{\mathbb{G}_{\mu\nu}\left(x\right)dx^{\mu}dx^{\nu}}}+r_{0}^{2}d\Omega_{n+1}^{2}.
\end{equation}

\noindent Based on this metric, it is straightforward to prove that
the action (\ref{eq:reduced action}), including complete $\alpha^{\prime}$
corrections, can be rewritten as the Hohm-Zwiebach action by following
these steps:
\begin{enumerate}
\item The closed string fields $g$ and $\phi_{D}$ in the action (\ref{eq:reduced action})
only depend on the coordinate $x$ as $n\rightarrow\infty$. According
to Sen's proof \cite{Sen:1991zi,Sen:1991cn}, this yields an emergent
$O\left(d,d\right)$ symmetry in the complete action at leading order
in $\frac{1}{n}$. 
\item Given that the radius of the metric above remains fixed at $r_{0}$,
the spherical coordinates can be integrated out. Consequently, the
action (\ref{eq:reduced action}) reduces to two dimensions, where
the original $O\left(d,d\right)$ symmetry simplifies to $O\left(1,1\right)$
symmetry. The corresponding two-dimensional action is $O\left(1,1\right)$
invariant.
\end{enumerate}
If we assume the two-dimensional metric takes the form:

\begin{equation}
ds^{2}=\underset{2\;\mathrm{dimensions}}{\underbrace{\mathbb{G}_{\mu\nu}\left(x\right)dx^{\mu}dx^{\nu}}}=-a\left(\rho\right)^{2}dt^{2}+b\left(\rho\right)^{2}d\rho^{2},
\end{equation}

\noindent then the action (\ref{eq:reduced action}) can be reformulated
as the two-dimensional $O\left(1,1\right)$ invariant Hohm-Zwiebach
action, as described in \cite{Wang:2019mwi}:

\begin{equation}
I_{\mathrm{String}}^{\left(\mathrm{all}\right)}=-\frac{\Omega_{n+1}r_{0}^{n+1}}{16\pi G_{D}}\int d^{2}xe^{-\Phi}\left(-\frac{1}{b}\Phi^{\prime2}-\sum_{\text{k=1}}^{\infty}\left(-\alpha^{\prime}\right)^{k-1}\frac{2^{2k+1}}{b^{2k-1}}\bar{c}_{k}H^{2k}+4\lambda^{2}\right)+\mathcal{O}\left(\frac{1}{n}\right).\label{eq:O(1,1) action}
\end{equation}

\noindent where $H\equiv\frac{\partial_{\rho}a\left(\rho\right)}{a\left(\rho\right)}$,
$\Phi^{\prime}\left(\rho\right)\equiv\partial_{\rho}\Phi\left(\rho\right)$,
and the coefficients are given by $\bar{c}_{1}=c_{1}=-\frac{1}{8}$,
$\bar{c}_{2}=-c_{2}=-\frac{1}{64}$, $\bar{c}_{2k-1}=c_{2k-1}$, and
$\bar{c}_{2k}=-c_{2k}$. The $O\left(1,1\right)$ invariant dilaton
is defined as:

\begin{equation}
\Phi\left(\rho\right)=2\phi\left(\rho\right)-\ln\sqrt{-\mathrm{det}\mathbb{G}_{00}}.
\end{equation}

\noindent It is important to note the relationship between the $O\left(1,1\right)$
invariant dilaton $\Phi\left(\rho\right)$ and the shifted dilaton
$\tilde{\phi}\left(\rho\right)$: $\Phi\left(\rho\right)=\tilde{\phi}\left(\rho\right)+\phi\left(\rho\right)$. 

In summary, as $n\rightarrow\infty$, the $D$-dimensional low-energy
effective action of the closed string, incorporating complete $O\left(1,1\right)$invariant
Hohm-Zwiebach action. This formulation provides a systematic method
for studying the $\alpha^{\prime}$-corrected near-horizon/singularity
geometries of black holes derived from both gravity and string theory.

\subsection{Near-horizon/singularity geometries with complete $\alpha^{\prime}$
correction}

\noindent To derive the EOM from the action (\ref{eq:O(1,1) action}),
we set $\frac{\Omega_{n+1}r_{0}^{n+1}}{16\pi G_{D}}=1$. The EOM are
expressed as:

\begin{eqnarray}
\frac{1}{b^{2}}\Phi^{\prime\prime}-\frac{b^{\prime}}{b^{3}}\Phi^{\prime}+\frac{1}{2}\frac{1}{b}Hf\left(H\right) & = & 0,\nonumber \\
\frac{d}{dr}\left(e^{-\Phi}f\left(H\right)\right) & = & 0,\nonumber \\
\frac{1}{b^{2}}\Phi^{\prime2}+g\left(H\right)-4\lambda^{2} & = & 0,\label{eq:corrected EOM}
\end{eqnarray}

\noindent with an additional constraint $bg^{\prime}\left(H\right)=Hf^{\prime}\left(H\right)$,
where:

\begin{eqnarray}
f\left(H\right) & = & \sum_{\text{k=1}}^{\infty}\left(-\alpha^{\prime}\right)^{k-1}2^{2\left(k+1\right)}k\bar{c}_{k}\left(\frac{1}{b}H\right)^{2k-1}=-2\left(\frac{1}{b}H\right)-128\bar{c}_{2}\left(\frac{1}{b}H\right)^{3}\alpha^{\prime}+\cdots,\nonumber \\
g\left(H\right) & = & \sum_{\text{k=1}}^{\infty}\left(-\alpha^{\prime}\right)^{k-1}2^{2k+1}\left(2k-1\right)\bar{c}_{k}\left(\frac{1}{b}H\right)^{2k}=-\left(\frac{1}{b}H\right)^{2}-96\bar{c}_{2}\left(\frac{1}{b}H\right)^{4}\alpha^{\prime}+\cdots.\nonumber \\
\end{eqnarray}

\noindent These EOM exhibit $O\left(1,1\right)$ symmetry:

\begin{equation}
\Phi\longleftrightarrow\Phi,\qquad H\longleftrightarrow-H.
\end{equation}

Our aim is to determine the non-perturbative and non-singular solutions
to the EOM (\ref{eq:corrected EOM}) incorporating complete $\alpha^{\prime}$
corrections. The strategy is outlined as follows:
\begin{enumerate}
\item \textbf{Calculate the perturbative solutions to the EOM} (\ref{eq:corrected EOM})
in an order-by-order manner with respect to $\alpha^{\prime}$. In
this paper, we only calculate the perturbative solutions up to the
first order of $\alpha^{\prime}$ correction as an example. Higher
orders can be obtained similarly.
\item \textbf{Determine non-perturbative and non-singular solutions} $\Phi$
and $H$, which cover the perturbative solution as $\alpha^{\prime}\rightarrow0$
in the first step and solve the EOM (\ref{eq:corrected EOM}).
\end{enumerate}
We can now proceed to calculate the perturbative solutions of the
EOM (\ref{eq:corrected EOM}). To do this, we introduce a new variable
$\Omega$, defined as
\begin{equation}
\Omega\equiv e^{-\Phi},\label{eq:pertur notation}
\end{equation}

\noindent where $\Omega^{\prime}=-\Phi^{\prime}\Omega$ and $\Omega^{\prime\prime}=\left(-\Phi^{\prime\prime}+\Phi^{\prime2}\right)\Omega$.
The EOM (\ref{eq:corrected EOM}) become

\noindent 
\begin{eqnarray}
\Omega^{\prime\prime}-\frac{b^{\prime}}{b}\Omega^{\prime}-\left(h\left(H\right)+4\lambda^{2}\right)\Omega b^{2} & = & 0,\nonumber \\
\frac{d}{dr}\left(\Omega f\left(H\right)\right) & = & 0,\nonumber \\
\Omega^{\prime2}+\left(g\left(H\right)-4\lambda^{2}\right)\Omega^{2}b^{2} & = & 0,\label{eq:reEOM}
\end{eqnarray}

\noindent where we define a new function

\begin{equation}
h\left(H\right)\equiv\frac{1}{2}\frac{1}{b}Hf\left(H\right)-g\left(H\right)=\bar{c}_{2}\left(\frac{1}{b}H\right)^{4}\alpha^{\prime}+\ldots,
\end{equation}

\noindent It is easy to see that $h\left(H\right)=0$ at the zeroth
order of $\alpha^{\prime}$. Now, we assume the perturbative solutions
of the EOM (\ref{eq:reEOM}) take the forms:

\begin{eqnarray}
\Omega\left(\rho\right) & = & \Omega_{0}\left(\rho\right)+\alpha^{\prime}\Omega_{1}\left(\rho\right)+\ldots,\nonumber \\
H\left(\rho\right) & = & H_{0}\left(\rho\right)+\alpha^{\prime}H_{1}\left(\rho\right)+\ldots.\nonumber \\
b\left(\rho\right) & = & b_{0}\left(\rho\right),
\end{eqnarray}

\noindent where $\Omega_{i}$ and $H_{i}$ denote the $i$-th order
of the perturbative solutions. Substituting these perturbative series
into the functions $h\left(H\right)$, $f\left(H\right)$, and $g\left(H\right)$,
we obtain:

\begin{eqnarray}
h\left(H\right) & = & 32\bar{c}_{2}\left(\frac{1}{b_{0}}H_{0}\right)^{4}\alpha^{\prime}+\ldots,\nonumber \\
f\left(H\right) & = & -2\left(\frac{1}{b_{0}}H_{0}\right)-2\left(64\bar{c}_{2}\left(\frac{1}{b_{0}}H_{0}\right)^{3}+\frac{1}{b_{0}}H_{1}\right)\alpha^{\prime}+\cdots,\nonumber \\
g\left(H\right) & = & -\left(\frac{1}{b_{0}}H_{0}\right)^{2}-2\left(48\bar{c}_{2}\left(\frac{1}{b_{0}}H_{0}\right)^{4}+\frac{1}{b_{0}^{2}}H_{0}H_{1}\right)\alpha^{\prime}+\cdots.
\end{eqnarray}

\noindent Based on these expansions, the EOM (\ref{eq:reEOM}) give
us the differential equations at each order of $\alpha^{\prime}$.
At the zeroth order of $\alpha^{\prime}$, the EOM are:

\begin{eqnarray}
\Omega_{0}^{\prime\prime}-\frac{b_{0}^{\prime}}{b_{0}}\Omega_{0}^{\prime}-4\lambda^{2}\Omega_{0}b_{0}^{2} & = & 0,\nonumber \\
\frac{d}{dr}\left(-2\Omega_{0}\frac{1}{b_{0}}H_{0}\right) & = & 0,\nonumber \\
\Omega_{0}^{\prime2}-\left(\frac{1}{b_{0}^{2}}H_{0}^{2}+4\lambda^{2}\right)\Omega_{0}^{2}b_{0}^{2} & = & 0.\label{eq:0th EOM-1}
\end{eqnarray}

\noindent The solution is

\begin{equation}
\Omega_{0}\left(x\right)=\sinh\left(2\lambda x\right),\qquad H_{0}\left(x\right)=\pm2\lambda\text{csch}\left(2\lambda x\right),\qquad b_{0}=1,
\end{equation}

\noindent where we have employed the coordinate transformations $\rho=\frac{n}{2r_{0}}x$
and $t=\frac{n}{2r_{0}}t$. Note that the $+$ sign relates to the
near-horizon metric (\ref{eq:2d metric1}) of Schwarzschild-Tangherlini
black hole, while the $-$ sign relates to the near-singularity metric
(\ref{eq:2d metric2}) of the string black hole. Given the invariance
of the EOM under $\Phi\longleftrightarrow\Phi$ and $H\longleftrightarrow-H$,
we can simultaneously calculate the $\alpha^{\prime}$ corrections
to both metrics. Next, we focus on the (\ref{eq:reEOM}) at the first
order of $\alpha^{\prime}$:

\begin{eqnarray}
\Omega_{1}^{\prime\prime}+\left(-32\bar{c}_{2}H_{0}^{4}\Omega_{0}-4\lambda^{2}\Omega_{1}\right) & = & 0,\nonumber \\
\left(-2H_{0}\Omega_{1}-2\Omega_{0}\left(64\bar{c}_{2}H_{0}^{3}+H_{1}\right)\right) & = & 0,\nonumber \\
2\Omega_{0}^{\prime}\Omega_{1}^{\prime}-2\Omega_{0}\Omega_{1}\left(H_{0}^{2}+4\lambda^{2}\right)-2H_{0}\Omega_{0}^{2}\left(48\bar{c}_{2}H_{0}^{3}+H_{1}\right) & = & 0.
\end{eqnarray}

\noindent The first-order solutions are:

\begin{equation}
\Omega_{1}\left(x\right)=-\lambda^{2}\frac{\cosh\left(4\lambda x\right)}{\sinh\left(2\lambda x\right)},\qquad H_{1}\left(x\right)=\pm2\lambda^{3}\frac{\left(\cosh\left(4\lambda x\right)+4\right)}{\sinh^{3}\left(2\lambda x\right)}.
\end{equation}

\noindent Thus, the perturbative solution up to first order in $\alpha^{\prime}$
is given by:

\begin{eqnarray}
H\left(x\right) & = & \pm2\lambda\text{csch}\left(2\lambda x\right)\pm2\lambda^{3}\frac{\left(\cosh\left(4\lambda x\right)+4\right)}{\sinh^{3}\left(2\lambda x\right)}\alpha^{\prime}+\cdots,\nonumber \\
\Omega\left(x\right) & = & \sinh\left(2\lambda x\right)-\lambda^{2}\frac{\cosh\left(4\lambda x\right)}{\sinh\left(2\lambda x\right)}\alpha^{\prime}+\cdots,\label{eq:perturbed solution}
\end{eqnarray}

\noindent From the perturbative solution and the expression for $\Phi\left(x\right)$:

\begin{equation}
\Phi\left(x\right)=-\ln\left(\sinh\left(2\lambda x\right)\right)+\lambda^{2}\left(\coth^{2}\left(2\lambda x\right)+1\right)\alpha^{\prime}+\cdots,\label{eq:perturbed solution dilaton}
\end{equation}

\noindent and the perturbative solutions (\ref{eq:perturbed solution})
and (\ref{eq:perturbed solution dilaton}), we can derive the non-perturbative
and non-singular solution after careful analysis:

\begin{equation}
ds^{2}=-a\left(x\right)^{2}dt^{2}+dx^{2},
\end{equation}

\noindent where

\begin{eqnarray}
\Phi\left(x\right) & = & \frac{1}{2}\ln\left(\text{csch}^{2}\left(2\lambda x\right)+2\alpha^{\prime}\lambda^{2}\cosh\left(4\lambda x\right)\text{csch}^{4}\left(2\lambda x\right)\right),\nonumber \\
H\left(x\right) & = & \pm\frac{\lambda\text{csch}^{6}\left(2\lambda x\right)}{4\left(2\alpha^{\prime}\lambda^{2}\cosh\left(4\lambda x\right)\text{csch}^{2}\left(2\lambda x\right)+1\right)^{2}\left(2\alpha^{\prime}\lambda^{2}\cosh\left(4\lambda x\right)\text{csch}^{4}\left(2\lambda x\right)+\text{csch}^{2}\left(2\lambda x\right)\right)^{1/2}}\nonumber \\
 &  & \times\left(-4\alpha^{\prime}\lambda^{2}+4\left(2\alpha^{\prime}\lambda^{2}-1\right)\left(4\alpha^{\prime}\lambda^{2}+1\right)\cosh\left(4\lambda x\right)+\left(4\alpha^{\prime}\lambda^{2}+1\right)\left(8\alpha^{\prime}\lambda^{2}+1\right)\cosh\left(8\lambda x\right)+3\right),\nonumber \\
f\left(x\right) & = & -4\lambda\left(2\alpha^{\prime}\lambda^{2}\cosh\left(4\lambda x\right)\text{csch}^{4}\left(2\lambda x\right)+\text{csch}^{2}\left(2\lambda x\right)\right)^{1/2},\nonumber \\
g\left(x\right) & = & 4\lambda^{2}\left(1-\frac{\coth^{2}\left(2\lambda x\right)\left(4\alpha^{\prime}\lambda^{2}+\left(4\alpha^{\prime}\lambda^{2}+1\right)\cosh\left(4\lambda x\right)-1\right)^{2}}{\left(\left(4\alpha^{\prime}\lambda^{2}+1\right)\cosh\left(4\lambda x\right)-1\right)^{2}}\right).\label{eq:solution}
\end{eqnarray}

\noindent They provide non-perturbative and non-singular solutions
for the near-horizon geometry of the Schwarzschild-Tangherlini black
hole and the near-singularity geometry of the string black hole in
the large $D$ limit. Based on the ansatz (\ref{eq:ansatz}), the
Kretschmann scalar is given by:

\begin{equation}
\mathbb{R}_{\mu\nu\rho\sigma}\mathbb{R}^{\mu\nu\rho\sigma}=2\mathbb{R}_{\mu\nu}\mathbb{R}^{\mu\nu}=\mathbb{R}^{2}=4\left(\dot{H}+H^{2}\right)^{2}.
\end{equation}

\noindent The explicit form of $\mathbb{R}$ (we calculate only the
+ sign of $H\left(x\right)$ for simplicity) is:

\begin{eqnarray}
\mathbb{R}\left(x\right) & = & \frac{1}{32\left(2\alpha^{\prime}\lambda^{2}\cosh\left(4\lambda x\right)\text{csch}^{2}\left(2\lambda x\right)+1\right)^{5}\left(2\alpha^{\prime}\lambda^{2}\cosh\left(4\lambda x\right)\text{csch}^{4}\left(2\lambda x\right)+\text{csch}^{2}\left(2\lambda x\right)\right)^{1/2}}\times\nonumber \\
 &  & \left[\lambda^{2}\text{csch}^{10}(2\lambda x)\left(-8\left(8\alpha^{\prime}\lambda^{2}\left(8\alpha^{\prime}\lambda^{2}\left(2\alpha^{\prime}\lambda^{2}\left(10\alpha^{\prime}\lambda^{2}+7\right)+5\right)+7\right)-5\right)\left(4\alpha^{\prime}\lambda^{2}+1\right)\right.\right.\nonumber \\
 &  & -16\left(8\alpha^{\prime}\lambda^{2}+1\right)\left(4\alpha^{\prime}\lambda^{2}+1\right)^{4}\cosh^{3}(4\lambda x)+16\left(8\alpha^{\prime}\lambda^{2}\left(20\alpha^{\prime}\lambda^{2}+7\right)-3\right)\times\nonumber \\
 &  & \left(4\alpha^{\prime}\lambda^{2}+1\right)^{2}\cosh\left(4\lambda x\right)-4\left(4\alpha^{\prime}\lambda^{2}+1\right)\cosh\left(8\lambda x\right)\left(2\left(4\alpha^{\prime}\lambda^{2}+1\right)^{2}\left(16\alpha^{\prime}\lambda^{2}\left(5\alpha^{\prime}\lambda^{2}+1\right)-3\right)\right.\nonumber \\
 &  & \left.-\sqrt{2}\left(4\alpha^{\prime}\lambda^{2}\left(4\alpha^{\prime}\lambda^{2}\left(4\alpha^{\prime}\lambda^{2}-5\right)+5\right)+7\right)\text{csch}\left(4\lambda x\right)\sqrt{\text{csch}^{4}\left(2\lambda x\right)\left(\left(4\alpha^{\prime}\lambda^{2}+1\right)\cosh\left(4\lambda x\right)-1\right)}\right)\nonumber \\
 &  & +\sqrt{2}\text{csch}\left(4\lambda x\right)\left(8\alpha^{\prime}\lambda^{2}\left(2\alpha^{\prime}\lambda^{2}\left(16\alpha^{\prime}\lambda^{2}\left(8\alpha^{\prime}\lambda^{2}+1\right)+3\right)+5\right)\right.\nonumber \\
 &  & +8\left(2\alpha^{\prime}\lambda^{2}-1\right)\left(4\alpha^{\prime}\lambda^{2}+1\right)\left(4\alpha^{\prime}\lambda^{2}\left(8\alpha^{\prime}\lambda^{2}+1\right)+7\right)\cosh\left(4\lambda x\right)\nonumber \\
 &  & +8\left(2\alpha^{\prime}\lambda^{2}-1\right)\left(4\alpha^{\prime}\lambda^{2}+1\right)^{2}\left(8\alpha^{\prime}\lambda^{2}+1\right)\cosh\left(12\lambda x\right)\nonumber \\
 &  & \left.+\left(4\alpha^{\prime}\lambda^{2}+1\right)^{2}\left(8\alpha^{\prime}\lambda^{2}+1\right)^{2}\cosh\left(16\lambda x\right)+35\right)\times\nonumber \\
 &  & \left.\left.\left.\sqrt{\text{csch}^{4}\left(2\lambda x\right)\left(\left(4\alpha^{\prime}\lambda^{2}+1\right)\cosh\left(4\lambda x\right)-1\right)}\right)\right)\right].
\end{eqnarray}

\begin{figure}[H]
\noindent \begin{centering}
\includegraphics[scale=0.7]{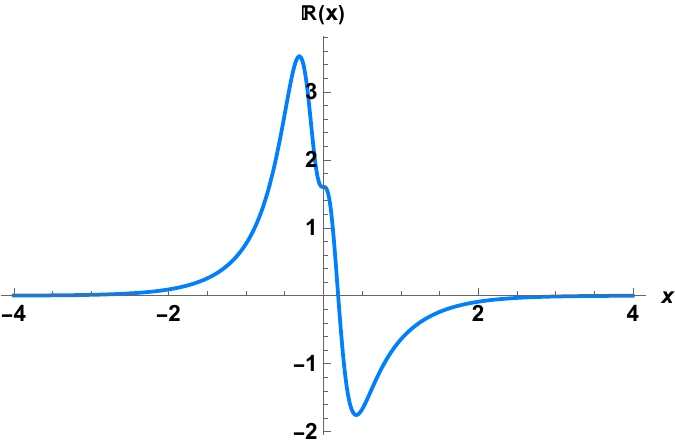}
\par\end{centering}
\caption{\label{fig:R}To plot the behavior of $\mathbb{R}\left(x\right)$,
we set $\alpha^{\prime}=10$ and $\lambda=1$.}
\end{figure}

\noindent We plot $\mathbb{R}\left(x\right)$ in figure (\ref{fig:R}),
demonstrating that it remains regular in the region $x\geq0$.

\section{Some useful results}

In this section, we present several types of black holes in the large
$D$ limit. These results can be applied to the Hohm-Zwiebach action
to include complete $\alpha^{\prime}$ corrections directly.

\subsection{(A)dS black hole}

Consider a $D=n+3$ dimensional low-energy effective action with a
constant dilaton $\phi_{0}$ and a positive cosmological constant:

\begin{equation}
I_{\mathrm{string}}=\frac{1}{16\pi G_{D}}\int d^{D}x\sqrt{-g}e^{-2\phi_{0}}\left(R-2\Lambda\right),\label{eq:dS action}
\end{equation}

\noindent where $\Lambda=\frac{\left(n+2\right)\left(n+1\right)}{2L^{2}}$.
The corresponding (A)dS black hole solution is given by:

\begin{equation}
ds^{2}=-f\left(r\right)dt^{2}+\frac{dr^{2}}{f\left(r\right)}+r^{2}d\Omega_{n+1}^{2},\qquad f\left(r\right)=1-\left(\frac{r_{0}}{r}\right)^{n}\mp\frac{r^{2}}{L^{2}},\label{eq:dS BH solution}
\end{equation}

\noindent where $-$ is used for dS, $+$ is used for AdS, and $L$
denotes the (A)dS radius.

To study the large $D$ limit of this geometry, we utilize the coordinate
transformation $\mathrm{R}=\left(\frac{r}{r_{0}}\right)^{n}$. With
this transformation, the metric (\ref{eq:dS BH solution}) becomes
\cite{Emparan:2013xia}:

\begin{equation}
ds^{2}=-\left(1-\frac{1}{\mathrm{R}}\mp\frac{r_{0}^{2}}{L^{2}}\mathrm{R}^{\frac{2}{n}}\right)dt^{2}+\frac{r_{0}^{2}}{n^{2}}\frac{\mathrm{R}^{\frac{2\left(1-n\right)}{n}}d\mathrm{R}^{2}}{\left(1-\frac{1}{\mathrm{R}}\mp\frac{r_{0}^{2}}{L^{2}}\mathrm{R}^{\frac{2}{n}}\right)}+r_{0}^{2}\mathrm{R}^{\frac{2}{n}}d\Omega_{n+1}^{2}.\label{eq:ADS metric}
\end{equation}

\noindent The event horizon is located at $\mathrm{R}=\mathrm{R}_{\mathrm{H}}$,
where

\begin{equation}
f\left(\mathrm{R}_{\mathrm{H}}\right)=1-\frac{1}{\mathrm{R}_{\mathrm{H}}}\mp\frac{r_{0}^{2}}{L^{2}}\mathrm{R}_{\mathrm{H}}^{\frac{2}{n}}=0.\label{eq:ads horizon}
\end{equation}

\noindent Taking the large $n$ limit where $\ln\mathrm{R}\ll n$,
we obtain:

\begin{equation}
ds^{2}=-\left(\frac{\mathrm{R}/\mathrm{R}_{0}-1}{\mathrm{R}}\right)dt^{2}+\frac{r_{0}^{2}}{n^{2}}\frac{d\mathrm{R}^{2}}{\mathrm{R}\left(\mathrm{R}/\mathrm{R}_{0}-1\right)}+r_{0}^{2}d\Omega_{n+1}^{2},
\end{equation}

\noindent where

\begin{equation}
\mathrm{R}_{0}=\frac{L^{2}}{L^{2}\mp r_{0}^{2}}.
\end{equation}

\noindent Rescaling $\left(t,\mathrm{R}\right)\rightarrow\left(\mathrm{R}_{0}t,\mathrm{R}_{0}\mathrm{R}\right)$,
the metric simplifies to:

\begin{equation}
ds^{2}=\mathrm{R}_{0}\left[-\left(\frac{\mathrm{R}-1}{\mathrm{R}}\right)dt^{2}+\frac{r_{0}^{2}}{n^{2}}\frac{d\mathrm{R}^{2}}{\mathrm{R}\left(\mathrm{R}-1\right)}\right]+r_{0}^{2}d\Omega_{n+1}^{2},
\end{equation}

\noindent Applying the following coordinate transformations:

\begin{equation}
\mathrm{R}=\cosh^{2}\rho,\qquad\left(\frac{2r_{0}}{n}\right)d\rho=d\rho,\qquad r_{0}^{2}d\Omega_{n+1}^{2}=\mathrm{R}_{0}^{2}d\Omega_{n+1}^{2},
\end{equation}

\noindent the metric becomes:

\begin{equation}
ds^{2}=\mathrm{R}_{0}\left(-\tanh^{2}\left(\frac{n}{2r_{0}}\rho\right)dt^{2}+d\rho^{2}\right)+\mathrm{R}_{0}^{2}d\Omega_{n+1}^{2},
\end{equation}

\noindent Thus, the difference between the large $D$ limit of the
(A)dS black hole and the Schwarzschild black hole is the shift of
the horizon from $r_{0}$ to $\mathrm{R}_{0}$. Moreover, $\mathrm{R}_{0}$
is precisely the event horizon as indicated by equation (\ref{eq:ads horizon})
in the large $D$ limit:

\begin{equation}
\mathrm{R}_{\mathrm{H}}=\mathrm{R}_{0}+\mathcal{O}\left(\frac{1}{n}\right).
\end{equation}

\noindent In the limit $L\rightarrow\infty$, this result reduces
to the Schwarzschild case.

Now, we wish to study a special case of the dS black hole when $r_{0}\sim L$.
The metric (\ref{eq:ADS metric}) simplifies to:

\begin{equation}
ds^{2}=\left(-\frac{1}{\mathrm{R}}\right)\left(-dt^{2}+\frac{r_{0}^{2}}{n^{2}}d\mathrm{R}^{2}\right)+r_{0}^{2}d\Omega_{n+1}^{2}.
\end{equation}

\noindent Given that $r_{0}^{n}=r_{\mathrm{H}}^{n}\left(1-\frac{2\Lambda}{\left(n+2\right)\left(n+1\right)}r_{\mathrm{H}}^{2}\right)$,
this provides a well-defined near-horizon geometry as $n\rightarrow\infty$.
Applying further coordinate transformations $2\frac{r_{0}}{n}\sqrt{R}=y$
and $2\frac{r_{0}}{n}dt=d\tau$, the metric becomes:

\begin{equation}
ds^{2}=\left(\frac{1}{y^{2}}d\tau^{2}-dy^{2}\right)+r_{0}^{2}d\Omega_{n+1}^{2}.\label{eq:dS metric solution}
\end{equation}

\noindent The two-dimensional geometry in the first bracket corresponds
to two-dimensional string cosmology.

Alternatively, from the perspective of the action, we obtain a consistent
result. Recalling the action (\ref{eq:dS action}), and performing
dimensional reduction on a sphere:

\begin{equation}
ds^{2}=\underset{2\;\mathrm{dimensions}}{\underbrace{\mathbb{G}_{\mu\nu}\left(x^{\mu}\right)dx^{\mu}dx^{\nu}}}+\underset{n+1\;\mathrm{dimensional\;sphere}}{\underbrace{r_{0}^{2}e^{-4\phi\left(x^{\mu}\right)/\left(n+1\right)}d\Omega_{n+1}^{2}}},
\end{equation}

\noindent the low-energy effective action with a positive cosmological
constant (\ref{eq:dS action}) becomes:

\begin{equation}
I_{\mathrm{string}}=\frac{\Omega_{n+1}r_{0}^{n+1}}{16\pi G_{D}}\int d^{2}x\sqrt{-\mathbb{G}}e^{-2\phi}\left(\mathbb{R}-2\Lambda+\frac{4n}{n+1}\left(\partial\phi\right)^{2}+\frac{n\left(n+1\right)}{r_{0}^{2}}e^{-4\phi/\left(n+1\right)}\right).
\end{equation}

\noindent In the limit $n\rightarrow\infty$, the action reduces to:

\begin{equation}
I_{EH}=\frac{1}{16\pi G_{2}}\int d^{2}x\sqrt{-\mathbb{G}}e^{-2\phi}\left(\mathbb{R}+4\left(\partial\phi\right)^{2}+\frac{n^{2}}{r_{0}^{2}}-\frac{n^{2}}{L^{2}}\right),
\end{equation}

\noindent where $G_{2}=\underset{n\rightarrow\infty}{\lim}\frac{G_{D}}{\Omega_{n+1}r_{0}^{n+1}}$.
Given the previous setting where $r_{0}\sim L$, the action finally
simplifies to:

\begin{equation}
I_{EH}=\frac{1}{16\pi G_{2}}\int d^{2}x\sqrt{-\mathbb{G}}e^{-2\phi}\left(\mathbb{R}+4\left(\partial\phi\right)^{2}\right),
\end{equation}

\noindent This results in the well-known string cosmology or domain
wall solution:

\begin{equation}
ds^{2}=\frac{1}{y^{2}}d\tau^{2}-dy^{2},
\end{equation}

\noindent which is consistent with (\ref{eq:dS metric solution})
from the perspective of the metric. The non-perturbative and non-singular
$\alpha^{\prime}$ corrected solution to this metric has been extensively
studied in the references \cite{Wang:2019kez,Wang:2019dcj,Ying:2021xse}. 

\subsection{AdS soliton}

The $D=n+3$ dimensional AdS soliton is a solution in an asymptotically
AdS space with an $S^{1}$ circle ($\tau\sim\tau+2\pi$), which can
be obtained through a double Wick rotation of the planar AdS solution
as described by Horowitz and Myers in \cite{Horowitz:1998ha}:

\begin{equation}
ds^{2}=\frac{r^{2}}{L^{2}}\left(-dt^{2}+\left(dx^{i}\right)_{n}^{2}+\left(1-\frac{r_{0}^{n+2}}{r^{n+2}}\right)d\tau^{2}\right)+\left(1-\frac{r_{0}^{n+2}}{r^{n+2}}\right)^{-1}\frac{L^{2}}{r^{2}}dr^{2},
\end{equation}

\noindent which can be seen as the gravitational dual of confining
gauge theory. Using the coordinate transformation $\mathrm{R}=\left(\frac{r}{r_{0}}\right)^{N}$
with $N\equiv n+2$, we get

\begin{equation}
ds^{2}=\frac{r_{0}^{2}}{L^{2}}\mathrm{R^{\frac{2}{N}}}\left(-dt^{2}+\left(dx^{i}\right)_{N-2}^{2}+\left(\frac{\mathrm{R}-1}{\mathrm{R}}\right)d\tau^{2}\right)+\frac{L^{2}}{N^{2}}\frac{dR^{2}}{\mathrm{R}\left(\mathrm{R}-1\right)}.
\end{equation}

\noindent Taking the large $D$ limit ($N\rightarrow\infty$), the
metric becomes:

\begin{equation}
ds^{2}=\frac{r_{0}^{2}}{L^{2}}\left(-dt^{2}+\left(\frac{\mathrm{R}-1}{\mathrm{R}}\right)d\tau^{2}\right)+\frac{L^{2}}{N^{2}}\frac{dR^{2}}{\mathrm{R}\left(\mathrm{R}-1\right)}+\frac{r_{0}^{2}}{L^{2}}\left(dx^{i}\right)_{N-2}^{2},
\end{equation}

\noindent where we need to keep $\frac{r_{0}^{2}}{L^{2}}$ and $\frac{L^{2}}{N^{2}}$
fixed. Using the coordinate transformations:

\begin{equation}
\mathrm{R}=\cosh^{2}\rho,\qquad\frac{L}{N}d\rho=\frac{r_{0}}{L}dy,
\end{equation}

\noindent the metric transforms to:

\begin{equation}
ds^{2}=\frac{r_{0}^{2}}{L^{2}}\left(-dt^{2}+\tanh^{2}\left(\frac{r_{0}N}{L^{2}}y\right)d\tau^{2}+dy^{2}\right)+\frac{r_{0}^{2}}{L^{2}}\left(dx^{i}\right)_{N-2}^{2}.
\end{equation}

\noindent The Hohm-Zwiebach action can also be directly applied to
this geometry based on the result of Section 4.2.

\subsection{Black branes}

Subsequently, one might wonder whether we can generalize this result
to more complicated backgrounds. As $n\rightarrow\infty$, the near-horizon
region of these backgrounds matches the geometries of higher-dimensional
string backgrounds, extending beyond just two-dimensional string black
hole. By employing the effective action approach instead of worldsheet
CFT approach, we can address this question.

To see it, let us consider the following $D=\left(d+1\right)+\left(n+1\right)$
dimensional neutral and non-extremal background:

\begin{equation}
ds^{2}=\underset{d+1\;\mathrm{dimensions}}{\underbrace{\mathbb{G}_{\mu\nu}\left(x\right)dx^{\mu}dx^{\nu}}}+\underset{n+1\;\mathrm{dimensional\;sphere}}{\underbrace{r_{0}^{2}e^{-4\phi\left(x\right)/\left(n+1\right)}d\Omega_{n+1}^{2}}},
\end{equation}

\noindent where $\mathbb{G}_{\mu\nu}\left(x\right)$ and $\phi\left(x\right)$
depend only on $x$. Substituting this ansatz into the $\alpha^{\prime}$
corrected low-energy effective action, and considering the large $D$
limit ($n\rightarrow\infty$), we obtain:
\begin{eqnarray}
I & = & \frac{1}{16\pi G_{D}}\int d^{D}x\sqrt{-g}e^{-2\phi_{D}}\left(R+4\left(\partial\phi_{D}\right)^{2}+\frac{1}{4}\alpha^{\prime}R^{\mu\nu\rho\sigma}R_{\mu\nu\rho\sigma}+\alpha^{\prime2}\left(\ldots\right)+\ldots\right)\nonumber \\
 & \stackrel{n\rightarrow\infty}{\rightarrow} & \frac{1}{16\pi G_{d+1}}\int d^{d+1}x\sqrt{-\mathbb{G}}e^{-2\tilde{\phi}}\left(\mathbb{R}+4\left(\partial\tilde{\phi}\right)^{2}+4\lambda^{2}+\frac{1}{4}\alpha^{\prime}\mathbb{R}^{\mu\nu\rho\sigma}\mathbb{R}_{\mu\nu\rho\sigma}+\alpha^{\prime2}\left(\ldots\right)+\ldots\right)+\mathcal{O}\left(\frac{1}{n}\right),\nonumber \\
\end{eqnarray}

\noindent where $\tilde{\phi}\left(x\right)\equiv\phi\left(x\right)+\phi_{D}\left(x\right)$,
$G_{d+1}=\underset{n\rightarrow\infty}{\lim}\frac{G}{\Omega_{n+1}r_{0}^{n+1}}$
and $\lambda=\frac{n}{2r_{0}}$. Now, since the metric of the action
only depends on one coordinate, and the action originates from the
low-energy effective action, this action possesses $O\left(d,d\right)$
symmetry. The corresponding metric becomes:

\begin{equation}
ds^{2}=\mathbb{G}_{\mu\nu}\left(x\right)dx^{\mu}dx^{\nu}+r_{0}^{2}d\Omega_{n+1}^{2},
\end{equation}

\noindent where the metric $\mathbb{G}_{\mu\nu}\left(x\right)$ can
always be transformed into the following form:

\begin{equation}
ds^{2}=-a_{1}\left(x\right)^{2}dt^{2}+a_{2}\left(x\right)^{2}dx^{2}+a_{3}\left(x\right)^{2}dy^{2}+a_{4}\left(x\right)^{2}dz^{2}+\ldots.\label{eq:set-up}
\end{equation}

\noindent Therefore, the action is nothing but the $d+1$dimensional
Hohm-Zwiebach action. Since the metric is anisotropic, the action
can be expressed as:

\begin{eqnarray}
I_{\mathrm{HZ}} & = & -\int dxa_{2}e^{-\Phi}\left(-\left(\mathcal{D}\Phi\right)^{2}+4\lambda^{2}+\sum_{\text{k=1}}^{\infty}\left(\alpha^{\prime}\right)^{k-1}\bar{c}_{k}\mathrm{Tr}\left(\left(\mathcal{D}\bar{\mathcal{S}}\right)^{2k}\right)+\mathrm{multi-traces}\right)\nonumber \\
 & = & -\int dxa_{2}e^{-\Phi}\left(-\left(\mathcal{D}\Phi\right)^{2}+\bar{c}_{1,0}\mathrm{Tr}\left(\left(\mathcal{D}\bar{\mathcal{S}}\right)^{2}\right)+4\lambda^{2}+\alpha^{\prime}\bar{c}_{2,0}\mathrm{Tr}\left(\left(\mathcal{D}\bar{\mathcal{S}}\right)^{4}\right)+\alpha^{\prime2}\bar{c}_{3,0}\mathrm{Tr}\left(\left(\mathcal{D}\bar{\mathcal{S}}\right)^{6}\right)\right.\nonumber \\
 &  & \left.+\alpha^{\prime3}\left[\bar{c}_{4,0}\mathrm{Tr}\left(\left(\mathcal{D}\bar{\mathcal{S}}\right)^{8}\right)+\bar{c}_{4,1}\left(\mathrm{Tr}\left(\left(\mathcal{D}\bar{\mathcal{S}}\right)^{4}\right)\right)^{2}\right]+\cdots\right),
\end{eqnarray}

\noindent with 

\begin{equation}
\sqrt{-\mathcal{G}}e^{-2\phi}=e^{-\Phi},\qquad\mathcal{D}\equiv\frac{1}{a_{2}\left(x\right)}\frac{\partial}{\partial x},\qquad\bar{\mathcal{S}}\equiv\left(\begin{array}{cc}
0 & \mathcal{G}_{ij}\\
\mathcal{G}_{ij}^{-1} & 0
\end{array}\right),\qquad\mathcal{G}_{ij}\equiv\left(\begin{array}{ccc}
-a_{1}\left(x\right)^{2} & 0 & 0\\
0 & a_{3}\left(x\right)^{2} & 0\\
0 & 0 & \ddots
\end{array}\right).
\end{equation}

\noindent where $\bar{c}_{1,0}=c_{1,0}=-\frac{1}{8}$, $\bar{c}_{2,0}=-c_{2,0}=-\frac{1}{64}$,
$\bar{c}_{3,0}=c_{3,0}=-\frac{1}{3\cdot2^{7}}$, $\bar{c}_{4,i}=-c_{4,i}$,
$\bar{c}_{2k-1,i}=c_{2k-1,i}$, $\bar{c}_{2k,i}=-c_{2k,i}$. Note
that there is no $a_{2}\left(x\right)^{2}$in the definition of $\mathcal{G}_{ij}$,
due to the definition of the $O\left(d,d\right)$ dilaton requiring:

\begin{equation}
\mathbb{G}_{\mu\nu}=\left(\begin{array}{cc}
a_{2}\left(x\right)^{2} & 0\\
0 & \mathcal{G}_{ij}
\end{array}\right).
\end{equation}

\noindent In our setup (\ref{eq:set-up}), since $\mathbb{G}_{\mu\nu}$
depends on $x$, $a_{2}\left(x\right)^{2}$ relates to a lapse-like
operator in the language of $O\left(d,d\right)$ formalism. The corresponding
EOM can be calculated directly, and are given as follows:

\begin{eqnarray}
\mathcal{D}^{2}\Phi+\frac{1}{2}\frac{1}{a_{2}}\stackrel[i=1]{d}{\sum}H_{i}f_{i}\left(H_{i}\right) & = & 0,\nonumber \\
\frac{d}{dx}\left(e^{-\Phi}f_{i}\left(H_{i}\right)\right) & = & 0,\nonumber \\
\left(\mathcal{D}\Phi\right)^{2}+\stackrel[i=1]{d}{\sum}g_{i}\left(H_{i}\right)-4\lambda^{2} & = & 0,
\end{eqnarray}

\noindent where $H_{i}\equiv\frac{\partial_{x}a_{i}\left(x\right)}{a_{i}\left(x\right)}$
and

\begin{eqnarray}
f_{i}\left(H_{i}\right) & = & -2\left(\frac{1}{a_{2}}H_{i}\right)+2\alpha^{\prime}\left(\frac{1}{a_{2}}H_{i}\right)^{3}-2\alpha^{\prime2}\left(\frac{1}{a_{2}}H_{i}\right)^{5},\nonumber \\
 &  & -8\alpha^{\prime3}4^{4}\left[2\bar{c}_{4,0}\left(\frac{1}{a_{2}}H_{i}\right)^{7}+4\bar{c}_{4,1}\left(\stackrel[j=1]{3}{\sum}\left(\frac{1}{a_{2}}H_{j}\right)^{4}\right)\left(\frac{1}{a_{2}}H_{i}\right)^{3}\right]+\cdots\nonumber \\
g_{i}\left(H_{i}\right) & = & -\left(\frac{1}{a_{2}}H_{i}\right)^{2}+\frac{3}{2}\alpha^{\prime}\left(\frac{1}{a_{2}}H_{i}\right)^{4}-\frac{5}{3}\alpha^{\prime2}\left(\frac{1}{a_{2}}H_{i}\right)^{6}\nonumber \\
 &  & -7\alpha^{\prime3}4^{4}\left[2\bar{c}_{4,0}\left(\frac{1}{a_{2}}H_{i}\right)^{8}+4\bar{c}_{4,1}\left(\stackrel[j=1]{3}{\sum}\left(\frac{1}{a_{2}}H_{j}\right)^{4}\right)\left(\frac{1}{a_{2}}H_{i}\right)^{4}\right]+\cdots.
\end{eqnarray}

\section{Discussion and conclusion}

In the first part of this paper, we computed the higher-dimensional
black hole solution of the low-energy effective action in string theory.
When the dilaton field is trivial, this solution reduces to the Schwarzschild-Tangherlini
black hole in gravity. Subsequently, we applied Buscher rules to derive
its T-dual black hole solution with a naked singularity. In the limit
of large $D$, the Schwarzschild-Tangherlini black hole simplifies
to a two-dimensional near-horizon geometry, while its T-dual black
hole solution transforms into a two-dimensional near-singularity geometry.
These solutions correspond to the two-dimensional low-energy effective
theory and are connected through scale-factor duality. In the second
part, we investigated large $D$ black holes within the framework
of the string effective action, including complete $\alpha^{\prime}$
corrections. As $D\rightarrow\infty$, the closed string massless
sectors become independent of the $d$ coordinates and thus possess
$O\left(d,d\right)$ symmetry. After integrating out the spherical
part, the $O\left(d,d\right)$ symmetry reduces to $O\left(1,1\right)$
symmetry. The two-dimensional action can then be described by the
Hohm-Zwiebach action. Using these results, we computed both perturbative
and non-perturbative solutions of this action and studied the $\alpha^{\prime}$-corrected
near-horizon/singularity geometries of higher-dimensional Schwarzschild-Tangherlini
and string black holes. Finally, we provided some well-known black
hole solutions that can be directly applied to the Hohm-Zwiebach action
to study their $\alpha^{\prime}$-corrected geometries.

Finally, we highlight the following key points:

\subsection{Black hole singularity}

At first, let us recall the Schwarzschild-Tangherlini solution in
$D=3+n$ dimensions

\begin{equation}
ds^{2}=-f\left(r\right)dt^{2}+\frac{dr^{2}}{f\left(r\right)}+r^{2}d\Omega_{n+1}^{2},\qquad f\left(r\right)=1-\left(\frac{r_{0}}{r}\right)^{n}.
\end{equation}

\noindent Applying the coordinate transformations

\begin{equation}
\left(\frac{r}{r_{0}}\right)^{n}=\cosh^{2}\rho,\qquad d\tau=\frac{n}{2r_{0}}dt,
\end{equation}

\noindent the metric transforms into:

\begin{equation}
ds^{2}=-\left(\frac{2r_{0}}{n}\right)^{2}\tanh^{2}\rho d\tau^{2}+\left(\frac{2r_{0}}{n}\right)^{2}\left(\cosh\rho\right)^{\frac{4}{n}}d\rho^{2}+r_{0}^{2}\left(\cosh\rho\right)^{\frac{4}{n}}d\Omega_{n+1}^{2}.
\end{equation}

\noindent To describe the black hole inside the event horizon, we
perform the analytic continuation $\rho\rightarrow i\rho$, yielding:

\begin{equation}
ds^{2}=\left(\frac{2r_{0}}{n}\right)^{2}\tan^{2}\rho d\tau^{2}-\left(\frac{2r_{0}}{n}\right)^{2}\left(\cos\rho\right)^{\frac{4}{n}}d\rho^{2}+r_{0}^{2}\left(\cos\rho\right)^{\frac{4}{n}}d\Omega_{n+1}^{2}.
\end{equation}

\noindent The corresponding Kretschmann scalar is given by:

\begin{equation}
R_{\mu\nu\rho\sigma}R^{\mu\nu\rho\sigma}=\frac{n\left(n+2\right)\left(n+1\right)^{2}}{r_{0}^{4}}\left(\frac{1}{\cos^{4}\rho}\right)^{\frac{n+2}{n}}.\label{eq:non limit R}
\end{equation}

\noindent Here, the event horizon is located at $\rho=0$,while the
curvature singularity is at $\rho=\frac{\pi}{2}$. Note that we have
not yet taken the large $D$ limit at this stage.

Next, we consider the near-horizon geometry of the large $D$ black
hole:

\begin{equation}
ds^{2}=\left(\frac{2r_{0}}{n}\right)^{2}\left(-\tanh^{2}\rho d\tau^{2}+d\rho^{2}\right)+r_{0}^{2}d\Omega_{n+1}.
\end{equation}

\noindent We also perform the analytic continuation $\rho\rightarrow i\rho$
and obtain:

\begin{equation}
ds^{2}=\left(\frac{2r_{0}}{n}\right)^{2}\left(-d\rho^{2}+\tan^{2}\left(\rho\right)d\tau^{2}\right)+r_{0}^{2}d\Omega_{n+1}.
\end{equation}

\noindent The Kretschmann scalar for the metric above is:

\begin{equation}
R_{\mu\nu\rho\sigma}R^{\mu\nu\rho\sigma}=\frac{n^{4}}{r_{0}^{4}}\left(\frac{1}{\cos^{4}\rho}\right),
\end{equation}

\noindent with the curvature singularity located at $\rho=\frac{\pi}{2}$.
Comparing this result with the original one (\ref{eq:non limit R}),
we observe that the large $D$ limit does not erase the singularity
information; both expressions feature the key function $\frac{1}{\cos^{4}\rho}$.
Therefore, it is reasonable to anticipate that the singularity can
be resolved through the near-horizon geometry using the two-dimensional
Hohm-Zwiebach action. This regular black hole solution has been given
in reference \cite{Ying:2022xaj}. Furthermore, in Section 3.2, we
demonstrate that the near-horizon geometry is dual to the near-singularity
geometry, with the singularity being resolved in Section 4.2. This
provides additional evidence for our hypothesis. We anticipate that
this result, together with our previous resolution of the Schwarzschild
black hole's singularity \cite{Wu:2024eci}, may ultimately address
the singularity problem in black holes.

\subsection{Lovelock gravity}

In Einstein's gravity, the most general extension of general relativity
to include higher curvature corrections is Lovelock gravity \cite{Lovelock:1971yv}.
Its corresponding Einstein tensor must satisfy the following three
criteria: 1) Symmetric; 2) Divergence free (Bianchi identity); 3)
Equations of motion (EOM) do not include higher derivatives (ghost-free).
As $D\rightarrow\infty$, the action of Lovelock gravity also has
an infinite series of correction terms, which makes the overall story
complicated. Therefore, it is worth to study the difference between
string's $\alpha^{\prime}$ correction and Lovelock extension for
large $D$ gravity.

For example, considering the Lovelock action which is constructed
from the dimensionally extended Euler densities in $D\left(\geq4\right)$
dimensions without matter sources \cite{Lovelock:1971yv}:

\begin{eqnarray}
I_{\mathrm{Lovelock}} & = & \int d^{D}x\sqrt{-g}\sum_{k=0}^{\left[\frac{D-1}{2}\right]}\alpha_{k}\mathcal{L}_{k},\nonumber \\
\mathcal{L}_{k} & \equiv & \frac{1}{2^{k}}\delta_{\rho_{1}\cdots\rho_{k}\sigma_{1}\cdots\sigma_{k}}^{\mu_{1}\cdots\mu_{k}\nu_{1}\cdots\nu_{k}}R_{\mu_{1}\nu_{1}}^{\quad\;\rho_{1}\sigma_{1}}\cdots R_{\mu_{k}\nu_{k}}^{\quad\;\rho_{k}\sigma_{k}},\label{eq:Love action}
\end{eqnarray}

\noindent where $\left[\left(D-1\right)/2\right]$ denotes the integer
part of $\left(D-1\right)/2$, $\delta_{\rho_{1}\cdots\rho_{k}\sigma_{1}\cdots\sigma_{k}}^{\mu_{1}\cdots\mu_{k}\nu_{1}\cdots\nu_{k}}$
is an anti-symmetric generalized Kronecker delta. Now, to consider
the background

\begin{equation}
ds^{2}=\underset{2\;\mathrm{dimensions}}{\underbrace{-a\left(x\right)^{2}dt^{2}+b\left(x\right)^{2}dx^{2}}}+\underset{n+1\;\mathrm{dimensional\;sphere}}{\underbrace{r_{0}^{2}e^{-4\phi\left(x^{\mu}\right)/\left(n+1\right)}d\Omega_{n+1}^{2}}},
\end{equation}

\noindent the Lovelock action becomes \cite{Kunstatter:2015vxa}:

\begin{equation}
I_{Lovelock\left(2D\right)}=\frac{r_{0}^{n+1}}{L^{n+1}}\int d^{2}xbe^{-\Phi}\sum_{k=0}^{\left[\frac{n+3}{2}\right]}\alpha_{k}\mathcal{L}_{k},
\end{equation}

\noindent with

\begin{eqnarray}
\mathcal{L}_{k} & = & \frac{\left(n+1\right)!}{\left(n-2k+3\right)!}\left(r_{0}a^{-\frac{1}{n+1}}e^{-\frac{\Phi}{n+1}}\right)^{-2k}\left[k\left(-2\mathcal{D}\mathcal{H}-2\mathcal{H}^{2}\right)\left(r_{0}a^{-\frac{1}{n+1}}e^{-\frac{\Phi}{n+1}}\right)^{2}\right.\nonumber \\
 &  & +\left(n-2k+3\right)\left(n-2k+2\right)\left\{ \left(1-\frac{r_{0}^{2}a^{-\frac{2}{n+1}}e^{-\frac{2\Phi}{n+1}}}{\left(n+1\right)^{2}}\left(\mathcal{D}\Phi+\mathcal{H}\right)^{2}\right)^{k}+2k\frac{r_{0}^{2}a^{-\frac{2}{n+1}}e^{-\frac{2\Phi}{n+1}}}{\left(n+1\right)^{2}}\left(\mathcal{D}\Phi+\mathcal{H}\right)^{2}\right\} \nonumber \\
 &  & +k\left(n-2k+3\right)r_{0}a^{-\frac{1}{n+1}}e^{-\frac{\Phi}{n+1}}\left\{ 1-\left(1-\frac{r_{0}^{2}a^{-\frac{2}{n+1}}e^{-\frac{2\Phi}{n+1}}}{\left(n+1\right)^{2}}\left(\mathcal{D}\Phi+\mathcal{H}\right)^{2}\right)^{k-1}\right\} \nonumber \\
 &  & \left.\times\left(-\frac{2}{n+1}r_{0}a^{-\frac{1}{n+1}}e^{-\frac{\Phi}{n+1}}\left(\mathcal{D}\left(\mathcal{D}\Phi+\mathcal{H}\right)-\frac{1}{\left(n+1\right)}\left(\mathcal{D}\Phi+\mathcal{H}\right)^{2}\right)\right)\right],
\end{eqnarray}

\noindent and

\begin{equation}
\sqrt{-\mathbb{G}}e^{-2\phi}=be^{-\Phi},
\end{equation}

\noindent where $\mathcal{D}f\left(x\right)\equiv\frac{1}{b}\partial_{x}f\left(x\right)$
and $\mathcal{H}=\frac{1}{b}\frac{\partial_{x}a}{a}$. Moreover, $b\left(x\right)$
can be seen as the lapse-like function in this ansatz. When $n\rightarrow\infty$,
and then keeping $\lambda=\frac{n}{2r_{0}}$ finite, we will have

\begin{eqnarray}
\mathcal{L}_{k} & = & 2k\left(\frac{n}{r_{0}}\right)^{2k-2}\left(\left(\mathcal{D}\Phi\right)^{2}+2\mathcal{H}\mathcal{D}\Phi-2\mathcal{D}\mathcal{H}-\mathcal{D}^{2}\Phi\right)\nonumber \\
 &  & +\left(\frac{n}{r_{0}}\right)^{2k}\left(1-\frac{r_{0}^{2}}{n^{2}}\left(\mathcal{D}\Phi+\mathcal{H}\right)^{2}+2k\frac{r_{0}^{2}}{n^{2}}\left(\mathcal{D}^{2}\Phi+\mathcal{D}\mathcal{H}\right)\right)\left(1-\frac{r_{0}^{2}}{n^{2}}\left(\mathcal{D}\Phi+\mathcal{H}\right)^{2}\right)^{k-1}+\mathcal{O}\left(\frac{1}{n}\right).\nonumber \\
\end{eqnarray}

\noindent It is interesting to study whether there exists a suitable
field redefinition such that all orders of the Lovelock extension
exhibit $O\left(1,1\right)$ symmetry in the large $D$ limit. If
the answer is affirmative, it would enable comparison with $\alpha^{\prime}$
corrections through such a field redefinition.

\noindent \bigskip 

\vspace{5mm}

\noindent {\bf Acknowledgements} 
We are deeply indebt to Houwen Wu for many discussions and suggestions. This work was supported by NSFC Grant No.12105031, No.12347101 and PSFC Grant No. cstc2021jcyj-bshX0227.

\end{document}